\documentclass[11pt,a4paper]{article}

\usepackage{amsmath,amssymb}
\usepackage{bm}

\def\Pexp{\mathop{\rm Pexp}\nolimits}

\def\rank{\mathop{\rm rank}\nolimits}

\newcommand{\CC}{\mathbb{C}}
\newcommand{\ZZ}{\mathbb{Z}}
\newcommand{\RR}{\mathbb{R}}
\newcommand{\ol}{\overline}
\newcommand{\olq}{{\tt q}}
\newcommand{\wh}{\widehat}

\newcommand{\rap}[2]
{\setbox1=\hbox{#1}%
\setbox2=\hbox to\wd1{\hss #2\hss}%
\mbox{\rlap{\box1}\box2}}

\usepackage{geometry}

\allowdisplaybreaks

\begin{document}

%titlepage
\begin{titlepage}
\title{
\vspace{-1.5cm}
\begin{flushright}
{\normalsize TIT/HEP-685\\ October 2021}
\end{flushright}
\vspace{1.5cm}
\LARGE{Holographic index calculation for
Argyres-Douglas and Minahan-Nemeschansky theories}}
\author{
Yosuke {\scshape Imamura\footnote{E-mail: imamura@phys.titech.ac.jp}},
and
Shuichi {\scshape Murayama.\footnote{E-mail: s.murayama@th.phys.titech.ac.jp}}
\\
\\
{\itshape Department of Physics, Tokyo Institute of Technology}, \\ {\itshape Tokyo 152-8551, Japan}}

\date{}
%\date{July 12, 2019}
\maketitle
\thispagestyle{empty}
\begin{abstract}
We calculate the superconformal indices of the $\mathcal N=2$ superconformal field theories
realized on $N$ coincident D3-branes in 7-brane backgrounds with constant axiodilaton via the AdS/CFT correspondence.
We include the finite-$N$ corrections as the contribution of D3-branes wrapped around 3-cycles in the internal space.
We take only single-wrapping contributions into account for simplicity.
We also determine the orders of the next-to-leading corrections which we do not calculate.
The orders are relatively high, and
we obtain many trustable terms.
We give the results for $N=1,2,3$ explicitly,
and find nice agreement with known results.
\end{abstract}

\end{titlepage}

\tableofcontents

%%%%%%%%%%%%%%%%%%%%%%%%%%%%%%%%%%%%%%%%%%%%%%%%%%%%%%%%%%%%%%%%%%%
\section{Introduction}

The recent development of quantum field theory owes a lot to brane realization in string theory and M-theory.
Various theories are realized as theories on branes placed in appropriate backgrounds.
Such theories are often strongly coupled, and many of them do not have known Lagrangian descriptions.
The brane realization provides non-perturbative methods to analyze such theories.
One of the most effective methods is the AdS/CFT correspondence \cite{Maldacena:1997re}.

If an SCFT is realized as the theory on $N$ coincident branes,
the brane system is well described by supergravity in the large $N$ limit, and
we can extract physical information about the SCFT by studying the classical solution.
For small $N$ the quantum gravity correction is expected to be important,
and in general qualitative analysis becomes difficult.
Even so, it has been proposed in \cite{Arai:2019xmp} that the finite $N$ correction to the
superconformal index \cite{Kinney:2005ej} can be calculated
as the contribution of giant gravitons
\cite{McGreevy:2000cw,Mikhailov:2000ya}
without taking account of quantum gravity.

In this paper we investigate four-dimensional ${\cal N}=2$ supersymmetric theories
realized on D3-branes put in
7-brane backgrounds with constant axiodilaton \cite{Sen:1996vd,Banks:1996nj,Dasgupta:1996ij,Douglas:1996js}.
We denote a theory in this class by $G[N]$,
where $G=H_0,H_1,H_2,D_4,E_6,E_7,E_8$ is the type of 7-brane
and $N$, which are called the rank of the theory, is the number of D3-branes.
$H_n[N]$ ($n=0,1,2$) are examples of a large class of theories called Argyres-Douglas theories \cite{Argyres:1995jj,Argyres:1995xn}
and $E_n[N]$ ($n=6,7,8$) are called Minahan-Nemeschansky theories \cite{Minahan:1996fg,Minahan:1996cj}.
$D_4[N]$ is an SQCD with the gauge group $Sp(N)$.

Let $X$, $Y$, and $Z$ be the three complex coordinates of $\CC^3$
transverse to the $N$ D3-branes.
The $7$-brane is placed at $Z=0$ (Table \ref{d3d7}).
\begin{table}[htb]
\caption{The brane setup.}\label{d3d7}
\centering
\begin{tabular}{ccccccccccc}
\hline
\hline
        & $0$ & $1$ & $2$ & $3$ & \multicolumn{2}{c}{$X$} & \multicolumn{2}{c}{$Y$} & \multicolumn{2}{c}{$Z$} \\
\hline
7-brane & $\circ$ & $\circ$ & $\circ$ & $\circ$ & $\circ$ & $\circ$ & $\circ$ & $\circ$ & $$ & $$ \\
D3-branes & $\circ$ & $\circ$ & $\circ$ & $\circ$ & $$ & $$ & $$ & $$ & $$ & $$ \\
\hline
\end{tabular}
\end{table}
The global symmetry of the SCFT is $SU(2,2|2)\times G\times SU(2)_F$,
and its maximal compact bosonic subgroup is
\begin{align}
U(1)_H\times SU(2)_J\times SU(2)_{\ol J}\times SU(2)_R\times U(1)_{R_Z}\times G\times SU(2)_F.
\end{align}
The subscripts of $U(1)$ and $SU(2)$ factors are generators of the factors.
$H$ is the Hamiltonian and $J$ and $\ol J$ are the angular momenta.
$U(1)_{R_Z}$ rotates the $Z$-plane, and
$SU(2)_R\times SU(2)_F$ rotates $\CC^2$ spanned by $X$ and $Y$.
We also define $R_X=R+F$ and $R_Y=R-F$ acting on
the $X$- and the $Y$-planes for later convenience.
$G$ is the gauge symmetry realized on the $7$-brane, and $G=H_0$, $H_1$, and $H_2$
are regarded as the trivial group, $SU(2)$, and $SU(3)$, respectively, as symmetry groups.
In addition, let $U(1)_A$ be the R-symmetry of the type IIB supergravity \cite{Schwarz:1983qr},
which is broken to its discrete subgroup due to the flux quantization.

The presence of the $7$-brane induces the deficit angle $\pi\alpha_G$ on the $Z$-plane
shown in Table \ref{deficit},
\begin{table}[htb]
\caption{The deficit angle parameters $\alpha_G$ of $7$-brane and
the dimensions $\Delta_G$ of Coulomb branch operators.}\label{deficit}
\centering
\begin{tabular}{cccccccc}
\hline
\hline
$G$        & $H_0$         & $H_1$         & $H_2$         & $D_4$ & $E_6$       & $E_7$         & $E_8$ \\
\hline
$\alpha_G$ & $\frac{1}{3}$ & $\frac{1}{2}$ & $\frac{2}{3}$ & $1$ & $\frac{4}{3}$ & $\frac{3}{2}$ & $\frac{5}{3}$ \\
$\Delta_G$ & $\frac{6}{5}$ & $\frac{4}{3}$ & $\frac{3}{2}$ & $2$ & $3$ & $4$ & $6$ \\
\hline
\end{tabular}
\end{table}
and the $Z$-plane is restricted by
\begin{align}
0\leq \arg Z\leq \pi(2-\alpha_G).
\label{restrictz}
\end{align}
The two boundary rays, $Z=r$ and $Z=re^{\pi i(2-\alpha_G)}$ ($r\in\RR_{\geq0}$),
are identified by the boundary condition
\begin{align}
{\cal O}(r)={\cal O}(e^{\pi i(2-\alpha_G)}r)
\equiv U_{\alpha_G}{\cal O}(r)U_{\alpha_G}^{-1},\quad
U_{\alpha_G}=e^{\pi i(2-\alpha_G)(R_Z-\frac{1}{2}A)},
\label{identification}
\end{align}
for an arbitrary local operator ${\cal O}(Z)$.
$R_Z$ and $A$ are normalized so that the supercharges
carry $R_Z=\pm\frac{1}{2}$ and $A=\pm1$,\footnote{Our normalization convention
for generators can be read off from (\ref{qcharges}).}
and the ${\cal N}=2$ supersymmetry is
generated by supercharges with $R_Z-\frac{1}{2}A=0$.
The globally defined coordinate $Z^{\Delta_G}$ corresponds to the
Coulomb branch operator with dimension $\Delta_G=2/(2-\alpha_G)$.
If $G=D_4,E_6,E_7,E_8$
then $\Delta_G$ is an integer,
and the identification (\ref{identification})
can be regarded as the orbifolding
by $\ZZ_{\Delta_G}$ generated by $U_{\alpha_G}$.

The system always contains a free hypermultiplet corresponding to
the ``center of mass'' degrees of freedom
along the $X$-$Y$ directions.
We exclude it from $G[N]$.
Then, if $N=1$, the $SU(2)_F$ becomes ineffective
and the flavor symmetry $G\times SU(2)_F$ reduces to $G$.
This fact will be useful in the analysis of rank $1$ theories
in section \ref{rank1.sec}.

The BPS operator spectrum of $G[\infty]$ is studied in \cite{Fayyazuddin:1998fb}
and \cite{Aharony:1998xz} by using the AdS/CFT correspondence.
The geometry is $AdS_5\times S^5_{\alpha_G}$ where $S_{\alpha_G}^5$
is the singular space defined from $S^5$ by the restriction
(\ref{restrictz}) and the identification (\ref{identification}).
The $7$-brane worldvolume is $AdS_5\times S^3$, where $S^3$ is the singular locus of
$S^5_{\alpha_G}$.
In the orbifold case considered in \cite{Fayyazuddin:1998fb}
the internal space is $S^5_{\alpha_G}=S^5/\ZZ_{\Delta_G}$,
and the Kaluza-Klein modes in the orbifold are
obtained from those in $S^5$ by
picking up $\ZZ_{\Delta_G}$ invariant modes,
whose quantum numbers satisfy
\begin{align}
R_z-\frac{1}{2}A\in \Delta_G\ZZ.
\label{kkquantiz}
\end{align}
The analysis in \cite{Fayyazuddin:1998fb}
was extended in \cite{Aharony:1998xz} in two ways:
The non-orbifold cases with $G=H_0,H_1,H_2$ were included by allowing fractional values of $\Delta_G$ in (\ref{kkquantiz}),
and fields living on the $7$-brane were taken into account
to generate the spectrum of operators transformed non-trivially under $G$.
The comparison with known results found nice agreement .

The purpose of this paper is to extend these analyses to $G[N]$ with finite $N$.
We use the superconformal index \cite{Kinney:2005ej}
\begin{align}
\mathcal I
&=\mathrm{tr}\left[e^{2\pi i(J+\ol J)} q^{H+\bar J}y^{2J}u_x^{R_X}u_y^{R_Y}u_z^{R_Z}\prod_{i=1}^{\rank G}x_i^{p_i}
\right]\quad(u_xu_yu_z=1)\nonumber\\
&=\mathrm{tr}\left[e^{2\pi i(J+\ol J)} q^{H+\bar J}y^{2J}
u_z^{R_Z-R}
u^{2F}
\prod_{i=1}^{\rank G}x_i^{p_i}
\right]\quad
\left(u=\sqrt{\frac{u_x}{u_y}}\right)
\label{scidef}
\end{align}
to express the BPS spectrum concisely, where
$p_i$ are Cartan generators of $G$.
This index is associated with the supercharge
${\cal Q}$ carrying the quantum numbers
\begin{align}
(H,J,\ol J,R_X,R_Y,R_Z,A)=(+\tfrac{1}{2},0,-\tfrac{1}{2},+\tfrac{1}{2},+\tfrac{1}{2},+\tfrac{1}{2},+1),
\label{qcharges}
\end{align}
and is contributed by operators saturating the corresponding bound
\begin{align}
\{{\cal Q},{\cal Q}^\dagger\}=H-2\ol J-R_X-R_Y-R_Z\geq0.
\label{bpsbound}
\end{align}

The large $N$ index is given on the AdS side by
\begin{align}
{\cal I}_{G}^{\rm KK}=\Pexp(i_G^{\rm grav}+i_G^{\rm vec}-i_{\rm hyp}).
\label{bulkcont}
\end{align}
The plethystic exponential $\Pexp$ is
defined by
\begin{align}
\Pexp
\sum_s c_sf_s
=\prod_s(1-f_s)^{-c_s},
\label{pexpdef}
\end{align}
where $f_s$ are monomials of fugacities and
$c_s$ are numerical coefficients.

$i_G^{\rm vec}$ is the single-particle index
of the gauge multiplet on the $7$-brane.
The mode expansion of the vector multiplet on the $7$-brane gives
the representation \cite{Aharony:1998xz}
\begin{align}
\bigoplus_{l=0}^\infty {\cal B}_{\frac{l+2}{2},0(0,0)}\otimes[\tfrac{l}{2}]_F\otimes R^G_\theta.
\end{align}
We denote the irreducible $G$-representation
with the highest weight $w$ by $R_w^G$,
and $R_\theta^G$ is the adjoint representation.
The corresponding single-particle index is
\begin{align}
i_G^{\rm vec}=\frac{q^2u_xu_y-q^3}{(1-q^{\frac{3}{2}}y)(1-q^{\frac{3}{2}}y^{-1})(1-qu_x)(1-qu_y)}\chi_\theta^G,
\label{id7}
\end{align}
where $\chi_\theta^G$ is the character of $R_\theta^G$.
The explicit form of $i_{\rm hyp}$, the contribution of the center-of-mass hypermultiplet, is
\begin{align}
i_{\rm hyp}
=\frac{q-q^2u_z}{(1-q^{\frac{3}{2}}y)(1-q^{\frac{3}{2}}y^{-1})}(u_x+u_y).
\label{ihyp}
\end{align}
$i_G^{\rm grav}$, the contribution of the gravity multiplet
in the bulk, will be calculated in the following sections.

For finite $N$, the index deviate from
(\ref{bulkcont})
at some order
of the $q$ expansion,
and we are interested in this finite $N$ correction.
We use the method used in \cite{Arai:2019xmp,Arai:2020qaj,Imamura:2021ytr},
which reproduces the finite $N$ correction to the index of ${\cal N}=4$ SYM as the
contribution of D3-branes wrapped on the internal space.
It has been also applied to more general theories \cite{Arai:2019wgv,Arai:2019aou,Arai:2020uwd,Fujiwara:2021xgu}.
For a large class of $4$-dimensional gauge theories
with the dual geometry $AdS_5\times M$
the index is given by
\begin{align}
{\cal I}={\cal I}^{\rm KK}\sum_{\vec n}{\cal I}_{\vec n},
\label{expansion}
\end{align}
where ${\cal I}^{\rm KK}$ is the index of massless fields
in $AdS\times M$,
and
is given by (\ref{bulkcont})
for the system of our interest.
${\cal I}_{\vec n}$ is the contribution
of wrapped D3-branes in the internal space $M$.
$\vec n=(n_1,\ldots,n_d)$ are wrapping numbers
around
appropriately chosen
supersymmetric cycles in $M$.
${\cal I}_{\vec n}$ is the
index of the brane system
specified by $\vec n$,
and is the product of two factors.
One is the classical factor ${\cal I}^{\rm cl}_{\vec n}$
determined by the energy and charges of the wrapped branes
without excitation of fields on them.
The other factor is the index of the field theory realized on the
brane system.
For $\vec n=\vec 0=(0,\ldots,0)$
there are no wrapped branes and ${\cal I}_{\vec 0}=1$.
In the large $N$ limit the other contributions decouple and the formula reduces
to the large $N$ relation
${\cal I}={\cal I}^{\rm KK}$,
while for finite $N$ ${\cal I}_{\vec n}$
with $\sum_In_I\geq1$ give finite $N$ corrections.

In the case of ${\cal N}=4$ $U(N)$ SYM
we use three three-cycles
in $S^5$ defined by $X=0$, $Y=0$, and $Z=0$,
and $\vec n=(n_x,n_y,n_z)$
\cite{Arai:2019xmp,Imamura:2021ytr}.
Although $S^5$ is replaced by $S^5_{\alpha_G}$ in the system of our interest,
we assume that the same choice of the three-cycles still works.

The classical factor is related to the volumes of the
cycles.
In the case of $AdS_5\times S^5$
it is given as follows.
Let $\Delta_{X=0}$
be the dimension of the operator corresponding to
a D3-brane wrapped on the cycle $X=0$.
$\Delta_{Y=0}$ and $\Delta_{Z=0}$ are also defined.
The RR flux $N$ and the $AdS_5$ radius $L$
are related by
$2\pi^2T_{\rm D3}L^4=N$.
According to the dictionary of the AdS/CFT correspondence,
this means $\Delta_{X=0}=\Delta_{Y=0}=\Delta_{Z=0}=N$,
and a wrapped brane
contributes $\sim q^N$ to the index.
If there are $n$ such branes, the corresponding term is $\sim q^{nN}$.
By taking account of the R-charges
carried by wrapped branes
we obtain the classical factor
${\cal I}^{\rm cl}_{(n_x,n_y,n_z)}=(qu_x)^{Nn_x}(qu_y)^{Nn_y}(qu_z)^{Nn_z}$.
If $S^5$ is replaced by $S^5_{\alpha_G}$
the volume
of the cycles are not all the same but are given by
\begin{align}
\Delta_{X=0}=\Delta_{Y=0}=N,\quad
\Delta_{Z=0}=\Delta_GN.
\end{align}
Correspondingly, the classical factor becomes
\begin{align}
{\cal I}^{\rm cl}_{(n_x,n_y,n_z)}=
(qu_x)^{n_xN}
(qu_y)^{n_yN}
(qu_z)^{n_z\Delta_GN}.
\label{clfact}
\end{align}
Because $\Delta_G>1$ for all $G$,
the leading correction is given by
${\cal I}_{(1,0,0)}$ and ${\cal I}_{(0,1,0)}$,
which are both of order $\sim q^N$.
In this paper we focus only on these two corrections for simplicity,
and calculate the approximate finite $N$ index by
\begin{align}
{\cal I}^{\rm AdS}_{G[N]}
={\cal I}^{\rm KK}_G(1+{\cal I}_{(1,0,0)}+{\cal I}_{(0,1,0)}).
\label{theformula}
\end{align}
We use the superscripts ``AdS'' to indicate the results are obtained by this equation.

(\ref{theformula}) is not exact but there must be an error
due to other contributions neglected in (\ref{theformula}).
\begin{align}
{\cal I}_{G[N]}={\cal I}_{G[N]}^{\rm AdS}+(\mbox{error}).
\label{adserr}
\end{align}
The source of the next-to-leading corrections
determining the order of the error term
in (\ref{adserr}) depends on $G$.
For $G=H_n$ the next-to-leading correction
is ${\cal I}_{(0,0,1)}\sim q^{\Delta_GN}$,
while for $G=E_n$
the next-to-leading corrections are
${\cal I}_{(2,0,0)}\sim{\cal I}_{(1,1,0)}\sim{\cal I}_{(0,2,0)}\sim q^{2N}$.
The $G=D_4$ case is marginal and all these contributions
give corrections of the same order.

In the order estimation above we
looked at only the classical factor (\ref{clfact})
and neglected the
contribution
of the field theory on the branes.
The latter often starts from a positive order term $\propto q^{\delta_G}$ ($\delta_G>0$),
and the order of ${\cal I}_{\vec n}$ is raised by $\delta_G$.
We will call this ``the tachyonic shift'' for the reason explained shortly.
With the tachyonic shift the expected order of the error term is given by
\begin{align}
H_n:{\cal O}(q^{\Delta_GN+\delta_G}),\quad
D_4, E_n:{\cal O}(q^{2N+\delta_G}).
\label{errororder}
\end{align}

Let us briefly explain how to calculate ${\cal I}_{(1,0,0)}$ and ${\cal I}_{(0,1,0)}$.
These two contributions are related by the Weyl reflection
$u_x\leftrightarrow u_y$,
and we consider ${\cal I}_{(1,0,0)}$,
the contribution of a D3-brane wrapped on $X=0$.
The theory realized on the brane is the supersymmetric $U(1)$ gauge theory
with the defect along the intersection with the $7$-brane.
The contribution of the vector multiplet
is given by $\Pexp i^{{\rm D3}(X=0)}_G$, where
the single-particle index
$i^{{\rm D3}(X=0)}_G$ can be calculated by
the mode analysis on the wrapped D3-brane.
The first few terms of the $q$-expansion of $i^{{\rm D3}(X=0)}_G$ is
\begin{align}
i^{{\rm D3}(X=0)}_G=\frac{1}{qu_x}+\frac{u_y}{u_x}+\cdots.
\label{id3qexp}
\end{align}
Detailed derivation will be given in the following sections.
See also Appendix \ref{d3.app}
The first term has the negative exponent of $q$,
and corresponds to the tachyonic mode with negative energy.
Such a mode exists because the cycle is topologically trivial
and the wrapped brane can be continuously unwrapped.
According to the definition (\ref{pexpdef})
the plethystic exponential is
\begin{align}
\Pexp i^{{\rm D3}(X=0)}_G
=\frac{1}{1-\frac{1}{qu_x}}\frac{1}{1-\frac{u_y}{u_x}}\cdots
=-\frac{qu_x}{1-qu_x}\frac{1}{1-\frac{u_y}{u_x}}\cdots.
\label{id3pexp}
\end{align}
The first factor corresponding to the tachyonic mode gives
the positive power of $q$.
This is the origin of the tachyonic shift in ${\cal I}_{(1,0,0)}$.
Similar shifts are expected in higher order contributions, too,
and $\delta_G$ in (\ref{errororder}) are the tachyonic shifts
in the next-to-leading corrections.

The defect contribution can be treated perturbatively only
in the $G=D_4$ case.
Then,
it will turns out that chiral fermions live on the defect and the Fock space
forms the basic representation of the affine $D_4$ algebra.
We calculate the defect contribution for $G\neq D_4$ on the assumption
that the Fock space of the defect degrees of freedom is the basic representation of
the affine $G$ algebra.

By combining the classical factor ${\cal I}_{(1,0,0)}=(qu_x)^N$,
the vector multiplet contribution (\ref{id3pexp}),
and the defect contribution $\chi^{\wh G}(qu_y)$
we obtain
\begin{align}
{\cal I}_{(1,0,0)}
&=(qu_x)^N\chi^{\wh G}(qu_y)\Pexp i^{{\rm D3}(X=0)}_G.
\label{i100}
\end{align}

This paper is organized as follows.
In the next section we confirm the method works well for the $D_4$ case.
Namely, we compare (\ref{theformula}) with the results on the gauge theory side
and confirm that the order of the error behaves like (\ref{errororder}).
Then, we move on to more interesting $G\neq D_4$ cases.
We analyze the rank $1$ theories $G[1]$ in Section \ref{rank1.sec}.
We find nice agreement with the known results in the literature,
and we determine $\delta_G$, which will be used in the
next section to determine the orders of errors for $N\geq2$.
In Section \ref{higherrank.sec} we calculate the indices for $N\geq 2$.
Again, the results are consistent with the known results.
Section \ref{disc.sec} is devoted to conclusions and discussion.
Appendices include technical details and some explicit results.

%%%%%%%%%%%%%%%%%%%%%%%%%%%%%%%%%%%%%%%%%%%%%%%%%%%%%%%
\section{$D_4[N]$}
\subsection{Large $N$ limit}
A $D_4$ $7$-brane is an O7-plane accompanied by four D7-branes \cite{Sen:1996vd}.
The field contents of the $Sp(N)$ gauge theory
realized on the D3-branes are shown in Table \ref{spntable}.
\begin{table}[htb]
\caption{Field contents of the $Sp(N)$ SQCD.
The anti-symmetric tensor representation
of $Sp(N)$ is not irreducible for $N\geq2$ but contains a singlet.
We exclude the singlet hypermultiplet from the
definition of $D_4[N]$.}\label{spntable}
\centering
\begin{tabular}{cccc}
\hline
\hline
       & $Sp(N)$ & $SU(2)_F$ & $D_4$ \\
\hline
vector & adj     & $\bm1$    & $\bm1$    \\
hyper  & anti-sym& $\bm2$    & $\bm1$    \\
hyper  & fund    & $\bm1$    & $\bm8_v$    \\
\hline
\end{tabular}
\end{table}
We can calculate the superconformal index on the gauge theory side
by using the localization formula for an arbitrary $N$.
For example, the result in the large $N$ limit is
\begin{align}
{\cal I}_{D_4[\infty]}=&
1
+q^2(u_z^2+\chi_2^F u_z^{-1} + u_z^{-1} \chi^{D_4}_{\bm{28}})
+q^{\frac{5}{2}}(-u_z \chi_1^J)
+\cdots,
\label{largenspn}
\end{align}
where $\chi_n^F$ and $\chi_n^J$ are $SU(2)_F$ and $SU(2)_J$ characters
\begin{align}
\chi_n^F=
u^n+u^{n-2}+
\cdots+u^{-n},\quad
\chi_n^J=y^n+y^{n-2}+\cdots+y^{-n},
\end{align}
and $\chi^G_{\bm{n}}$ is the character
of the $n$-dimensional irreducible $G$ representation.

It is easy to reproduce the large $N$ index
(\ref{largenspn}) by using (\ref{bulkcont}) on the AdS side.
$i^{\rm vect}_{D_4}$ and $i^{\rm hyp}$ are given in (\ref{id7}) and (\ref{ihyp}),
and $i^{\rm grav}_{D_4}$ are obtained as follows.

The supergravity Kaluza-Klein modes in $AdS_5\times S^5$
forms a series of $psu(2,2|4)$ representations ${\cal B}_{[0,n,0](0,0)}^{\frac{1}{2},\frac{1}{2}}$
($n=1,2,3,\ldots$) \cite{Gunaydin:1984fk,Kim:1985ez}.\footnote{We use the notation in \cite{Dolan:2002zh}
for ${\cal N}=2$ and ${\cal N}=4$ superconformal representations.}
After the supersymmetry is broken to ${\cal N}=2$ by the $7$-brane,
these are decomposed into
irreducible $su(2,2|2)\times SU(2)_F$ representations
${\cal B}_{\frac{m}{2},r(0,0)}\otimes[\frac{m}{2}]_F$ and ${\cal C}_{\frac{m}{2},r(0,0)}\otimes[\frac{m}{2}]_F$,
where $[s]_F$ is the spin $s$ $SU(2)_F$ representation
and $m$ and $r$ are integers.
(See Appendix \ref{grav.app}.)
We need to pick up $\ZZ_2$ invariant ones from them.
The components in these representations
carry
\begin{align}
R_z-\frac{1}{2}A=r,
\label{bulkrz}
\end{align}
and the orientifold projection leaves only representations with $r\in2\ZZ$.
By summing up their contributions we obtain
\begin{align}
i_{D_4}^{\rm grav}
&=\sum_{n=0}^\infty\sum_m i({\cal B}_{\frac{m}{2},2n(0,0)})\chi^F_m
+\sum_{n=0}^\infty\sum_m i({\cal C}_{\frac{m}{2},2n(0,0)})\chi^F_m
\nonumber\\
&=\frac{1}{2}\left(
\frac{qu_x}{1-qu_x}
+\frac{qu_y}{1-qu_y}
+\frac{qu_z}{1-qu_z}
-\frac{q^{\frac{3}{2}}y}{1-q^{\frac{3}{2}}y}
-\frac{q^{\frac{3}{2}}y^{-1}}{1-q^{\frac{3}{2}}y^{-1}}\right)
\nonumber\\
&+\frac{1}{4}\left(
\frac{(1+qu_x)(1+qu_y)(1-qu_z)(1+q^{\frac{3}{2}}y)(1+q^{\frac{3}{2}}y^{-1})}
{(1-qu_x)(1-qu_y)(1+qu_z)(1-q^{\frac{3}{2}}y)(1-q^{\frac{3}{2}}y^{-1})}-1\right).
\label{ibulkori}
\end{align}

${\cal I}^{\rm KK}_{D_4}$ obtained by substituting (\ref{ibulkori}), (\ref{id7}), and (\ref{ihyp})
to (\ref{bulkcont}) is shown in (\ref{ikkd4}), and it agrees with (\ref{largenspn}).

\newcommand{\eqo}{{\,\stackrel{\circ}{=}\,}}
%%%%%%%%%%%%%%%%%%%%%%%%%%%%%%%%%%%%%%%%%%%%%%%%%%%%%%%%%
\subsection{Finite $N$ corrections}
Let us consider the finite $N$ case.
In the following we often set all fugacities except for $q$ to be $1$ to save the space.
We will use ``$\eqo$'' to express the unrefinement.
For example,
the Kaluza-Klein index (\ref{ikkd4}), which is identical
with (\ref{largenspn}), is
expressed as follows.
\begin{align}
{\cal I}^{\rm KK}_{D_4}
\eqo
1+32q^2-2q^{\frac{5}{2}}+31q^3+62q^{\frac{7}{2}}+556q^4-4q^{\frac{9}{2}}+1117q^5+\cdots.
\end{align}
Let us compare this with the results for $N=1,2,3$
calculated on the gauge theory side:
\begin{align}
\mathcal I_{D_4[1]}\eqo&1+29q^2-2q^{\frac{5}{2}}-28q^3+60q^{\frac{7}{2}}+298q^4-60q^{\frac{9}{2}}-587q^5+\cdots\nonumber\\
\mathcal I_{D_4[2]}\eqo&1+32q^2-2q^{\frac{5}{2}}+27q^3+62q^{\frac{7}{2}}+467q^4-6q^{\frac{9}{2}}+632q^5+\cdots\nonumber\\
\mathcal I_{D_4[3]}\eqo&1+32q^2-2q^{\frac{5}{2}}+31q^3+62q^{\frac{7}{2}}+551q^4-4q^{\frac{9}{2}}+998q^5+\cdots.
\end{align}
We find that the finite $N$ corrections start at $q^{N+1}$.
\begin{align}
{\cal I}_{D_4[1]}-{\cal I}_{D_4}^{\rm KK}
&\eqo
-3q^2
-59q^3
-2q^{\frac{7}{2}}
-258q^4
-56q^{\frac{9}{2}}
-1704q^5
-566q^{\frac{11}{2}}
+\cdots,\nonumber\\
{\cal I}_{D_4[2]}-{\cal I}_{D_4}^{\rm KK}
&\eqo
-4q^3
-89q^4
-2q^{\frac{9}{2}}
-485q^5
-54q^{\frac{11}{2}}
-3671q^6
-588q^{\frac{13}{2}}
+\cdots,\nonumber\\
{\cal I}_{D_4[3]}-{\cal I}_{D_4}^{\rm KK}
&\eqo
-5q^4
-119q^5
-2q^{\frac{11}{2}}
-712q^6
-52q^{\frac{13}{2}}
-5648q^7
-590q^{\frac{15}{2}}
+\cdots.
\label{error1}
\end{align}
The refined expression for the leading terms in (\ref{error1}) is
\begin{align}
{\cal I}_{D_4[1]}-{\cal I}_{D_4}^{\rm KK}
&=
-q^{N+1}u_z^{-\frac{N+1}{2}}\chi^F_{N+1}+\cdots.
\label{d3leading}
\end{align}

%%%%%%%%%%%%%%%%%%%%%%%%%%%%%%%%%%%%%%%%%%%%%%%%%%%%%%%%%%%%%%%%%%%

We want to reproduce this finite $N$ correction
as the contribution of wrapped D3-branes.
Let us first consider ${\cal I}_{(1,0,0)}$,
the contribution of a D3-brane on $X=0$.
We have to consider fields arising from
two kinds of strings:
3-3 strings and 3-7 strings.

3-3 strings give an ${\cal N}=4$ vector multiplet
on the wrapped D3-brane.
Its fluctuation modes
belong to two series of representations
of unbroken supersymmetry.
Due to a similarity to the bulk modes
we use similar notation
${\cal B}^{{\rm D3}(X=0)}_{m,r}$ and ${\cal C}^{{\rm D3}(X=0)}_{m,r}$
for these representations.
See Appendix \ref{d3.app} for details.
The values of $R_Z-\frac{1}{2}A$ carried by the components of
these representations are
\begin{align}
R_Z-\frac{1}{2}A=r.
\end{align}
We obtain $i_{D_4}^{{\rm D3}(X=0)}$
by summing up all contributions from
${\cal B}^{{\rm D3}(X=0)}_{m,r}$ and ${\cal C}^{{\rm D3}(X=0)}_{m,r}$
with $r\in2\ZZ$:
\begin{align}
i^{{\rm D3}(X=0)}_{D_4}
&=\sum_{n=0}^\infty\sum_m i({\cal B}^{{\rm D3}(X=0)}_{m,2n})+\sum_{n=0}^\infty\sum_m i({\cal C}^{{\rm D3}(X=0)}_{m,2n})
\nonumber\\
&=\frac{1}{2}\left(\frac{(1+q^{-1}u_x^{-1})(1+q^{\frac{3}{2}}y)(1+q^{\frac{3}{2}}y^{-1})}{(1+qu_z)(1-qu_y)}
-\frac{(1-q^{-1}u_x^{-1})(1-q^{\frac{3}{2}}y)(1-q^{\frac{3}{2}}y^{-1})}{(1-qu_z)(1-qu_y)}\right).
\label{ix033}
\end{align}
The $q$-expansion of (\ref{ix033})
and its plethystic exponential have the form (\ref{id3qexp})
and (\ref{id3pexp}), respectively.

We also have the contribution from $3$-$7$ strings.
Because there are eight DN directions only chiral fermions
appear on the intersection.
They couple to the gauge symmetry on the D3-brane
and the $D_4$ symmetry on the 7-brane.
Important fact is that the $U(1)$ gauge symmetry
on the D3-brane is
broken to $\ZZ_2$ along the intersection
with O7-plane just like the gauge symmetry on a
type I D-string.
The $\ZZ_2$ gauged fermion system is nothing but the
free field realization of the $\wh D_4$ current algebra,
and the contribution to the index is
the character of the basic representation:
\begin{align}
\chi^{\wh D_4}(qu_y)
=\frac{1}{2}(Z_{\zeta=1}+Z_{\zeta=-1}),\quad
Z_\zeta
\equiv
\Pexp\left(-\frac{(qu_y)^{\frac{1}{2}}}{1-qu_y}\zeta\chi_{\bm{8}_v}^{D_4}\right).
\label{d4chiral}
\end{align}
Remark that we adopted the anti-periodic boundary condition for the fermions.
This is necessary to obtain the triality invariant spectrum required by the S-duality invariance of the
index \cite{Seiberg:1994aj,Lee:1997fy}.

(\ref{i100}) with 
(\ref{ix033}) and (\ref{d4chiral}) gives
\begin{align}
{\cal I}_{(1,0,0)}
=-\frac{(qu_x)^{N+1}}{1-\frac{u_y}{u_x}}
\left(1+
q(u_x + u_z^2u_x^{-1}-u_z+u_y^2u_x^{-1}+u_y \chi^{D_4}_{\bm{28}})
+\cdots
\right).
\label{ix0}
\end{align}
The contribution of a D3-brane around the $Y=0$ cycle, ${\cal I}_{(0,1,0)}$,
is obtained from this
by the Weyl reflection $u_x\leftrightarrow u_y$.
We can easily see that
the leading term (\ref{d3leading}) is reproduced:
\begin{align}
-\frac{(qu_x)^{N+1}}{1-\frac{u_y}{u_x}}
-\frac{(qu_y)^{N+1}}{1-\frac{u_x}{u_y}}
&=
-q^{N+1}u_z^{-\frac{N+1}{2}}\chi^F_{N+1}.
\label{theleadoing}
\end{align}
It will turn out that this term exists not only for $G=D_4$ but also for all $G$.
Some higher order terms are also correctly reproduced,
and the results of numerical calculation are
\begin{align}
{\cal I}_{D_4[1]}-{\cal I}_{D_4[1]}^{\rm AdS}
&\eqo -10q^5+20q^{\frac{11}{2}}-2124q^6+2028q^{\frac{13}{2}}-28273q^7+22214q^{\frac{15}{2}}+\cdots,
\nonumber\\
{\cal I}_{D_4[2]}-{\cal I}_{D_4[2]}^{\rm AdS}
&\eqo -20q^7+40q^{\frac{15}{2}}+\cdots,
\nonumber\\
{\cal I}_{D_4[3]}-{\cal I}_{D_4[3]}^{\rm AdS}
&\eqo -35q^9+\cdots.
\label{error2}
\end{align}
The order of these errors agree with (\ref{errororder}) with $\delta_{D_4}=3$,
and this suggests that the method works well
for $D_4[N]$.

%%%%%%%%%%%%%%%%%%%%%%%%%%%%%%%%%%%%%%%%%%%%%%%%%%%%%%%%%%%%%%%%%%%%%%%
\section{Rank one theories}\label{rank1.sec}
Now, let us apply our method to more interesting cases with $G\neq D_4$.
The Kaluza-Klein contribution (\ref{bulkcont}) is again calculated by using $i_G^{\rm vec}$ in
(\ref{id7}), $i_{\rm hyp}$ in (\ref{ihyp}),
and $i_G^{\rm grav}$ given by
\begin{align}
i_G^{\rm grav}
&=\sum_{n=0}^\infty\sum_m i({\cal B}_{\frac{m}{2},\Delta_G n(0,0)})\chi^F_m
+\sum_{n=0}^\infty\sum_m i(~{\cal C}_{\frac{m}{2},\Delta_G n(0,0)})\chi^F_m.
\label{iforzp}
\end{align}
The results for ${\cal I}_G^{\rm KK}$ are shown explicitly in Appendix \ref{largen.app}.

${\cal I}_{(1,0,0)}$ is calculated by (\ref{i100})
with the single-particle index
\begin{align}
i_G^{\rm D3}
&=\sum_{n=0}^\infty\sum_m i({\cal B}^{{\rm D3}(X=0)}_{\frac{m}{2},\Delta_G n})
+\sum_{n=0}^\infty\sum_m i({\cal C}^{{\rm D3}(X=0)}_{\frac{m}{2},\Delta_G n}).
\end{align}
Concerning the defect contribution, we assume that
it is the character $\chi^{\wh G}$ of the
basic representation of affine algebra $\wh G$
based on the result in the $D_4$ case.
(See Appendix \ref{affine.app} for the explicit form of $\chi^{\wh G}$.)
We obtain the following results.
\begin{subequations}
\label{sciresults}
\begin{align}
{\cal I}_{H_0[1]}^{\rm AdS}
&=1
+u_z^{\frac{6}{5}}q^{\frac{6}{5}}
-u_z^{\frac{1}{5}}\chi^J_1q^{\frac{17}{10}}
+u_z^{-\frac{4}{5}}q^{\frac{11}{5}}
+u_z^{\frac{12}{5}}q^{\frac{12}{5}}
+u_z^{\frac{6}{5}}\chi^J_1q^{\frac{27}{10}}
\underline{+u_z^{\frac{29}{5}}q^{\frac{14}{5}}}
+\cdots,\\
%%%%%%%%%%%%%%%%%%%%%%%%%%%%%%%%%%%%%%%%%
{\cal I}_{H_1[1]}^{\rm AdS}
&=1
+u_z^{\frac{4}{3}}q^{\frac{4}{3}}
-u_z^{\frac{1}{3}}\chi^J_1q^{\frac{11}{6}}
+u_z^{-1}\chi^{H_1}_{\bm{3}}q^2
+u_z^{-\frac{2}{3}}q^{\frac{7}{3}}
+u_z^{\frac{8}{3}}q^{\frac{8}{3}}
+u_z^{\frac{4}{3}}\chi^J_1q^{\frac{17}{6}}
-(1+\chi^{H_1}_{\bm{3}})q^3
\nonumber\\&
-u_z^{\frac{5}{3}}\chi^J_1q^{\frac{19}{6}}
+(-u_z^{\frac{1}{3}}\underline{+u_z^{\frac{19}{3}}}
     -u_z^{\frac{1}{3}}\chi^J_2)q^{\frac{10}{3}}
+\cdots,\\
%%%%%%%%%%%%%%%%%%%%%%%%%%%%%%%%%%%%%%%%%
{\cal I}_{H_2[1]}^{\rm AdS}
&=1
+u_z^{\frac{3}{2}}q^{\frac{3}{2}}
+(-u_z^{\frac{1}{2}}\chi^J_1+u_z^{-1}\chi^{H_2}_{\bm{8}})q^2
+u_z^{-\frac{1}{2}}q^{\frac{5}{2}}
+(-1+u_z^3+u_z^{\frac{3}{2}}\chi^J_1-\chi^{H_2}_{\bm{8}})q^3
\nonumber\\&
+(-u_z^{\frac{1}{2}}+u_z^{-1}\chi^J_1-u_z^2\chi^J_1
     -u_z^{\frac{1}{2}}\chi^J_2+u_z^{-1}\chi^{H_2}_{\bm{8}}\chi^J_1)q^{\frac{7}{2}}
+(2u_z+u_z^4\underline{+u_z^7}+u_z^{-\frac{1}{2}}\chi^J_1
\nonumber\\&
      +u_z^{-2}\chi^{H_2}_{\bm{27}})q^4
+\cdots,\\
%%%%%%%%%%%%%%%%%%%%%%%%%%%%%%%%%%%%%%%%%
{\cal I}_{D_4[1]}^{\rm AdS}
&=1
+(u_z^2+u_z^{-1}\chi^{D_4}_{\bm{28}})q^2
-u_z\chi^J_1q^{\frac{5}{2}}
-\chi^{D_4}_{\bm{28}}q^3
+(u_z^{-1}+u_z^2+u_z^{-1}\chi^{D_4}_{\bm{28}})\chi^J_1q^{\frac{7}{2}}
+(u_z^4
\nonumber\\&
       -u_z\chi^J_2+u_z^{-2}\chi^{D_4}_{\bm{300}})q^4
+(-1-u_z^3-\chi^{D_4}_{\bm{28}})\chi^J_1q^{\frac{9}{2}}
+(2u_z^{-1}\underline{+u_z^{-1}\chi^F_4+u_z^2\chi^F_2}+u_z^5
\nonumber\\&
    +u_z^{-1}\chi^J_2
    +u_z^2\chi^J_2-u_z^{-1}\chi^{D_4}_{\bm{28}}+u_z^{-1}\chi^J_2\chi^{D_4}_{\bm{28}}
    -u_z^{-1}\chi^{D_4}_{\bm{300}}-u_z^{-1}\chi^{D_4}_{\bm{350}})q^5
+\cdots,\\
%%%%%%%%%%%%%%%%%%%%%%%%%%%%%%%%%%%%%%%%%
{\cal I}_{E_6[1]}^{\rm AdS}&=1
+u_z^{-1}\chi^{E_6}_{\bm{78}}q^2
+(-1+u_z^3-\chi^{E_6}_{\bm{78}})q^3
+(u_z^{-1}-u_z^2+u_z^{-1}\chi^{E_6}_{\bm{78}})\chi^J_1q^{\frac{7}{2}}
+(2u_z
\nonumber\\&
        +u_z^{-2}\chi^{E_6}_{\bm{2430}})q^4
+(-2+u_z^3-\chi^{E_6}_{\bm{78}})\chi^J_1q^{\frac{9}{2}}
+(2u_z^{-1}\underline{+u_z^{-1}\chi^F_4}-u_z^2+u_z^{-1}\chi^J_2
\nonumber\\&
    -u_z^2\chi^J_2-u_z^{-1}\chi^{E_6}_{\bm{78}}
    +u_z^{-1}\chi^J_2\chi^{E_6}_{\bm{78}}
    -u_z^{-1}\chi^{E_6}_{\bm{2430}}
    -u_z^{-1}\chi^{E_6}_{\bm{2925}})q^5
+\cdots,\\
%%%%%%%%%%%%%%%%%%%%%%%%%%%%%%%%%%%%%%%%%
{\cal I}_{E_7[1]}^{\rm AdS}
&=1
+u_z^{-1}\chi^{E_7}_{\bm{133}}q^2
+(-1-\chi^{E_7}_{\bm{133}})q^3
+(u_z^{-1}+u_z^{-1}\chi^{E_7}_{\bm{133}})\chi^J_1q^{\frac{7}{2}}
+(u_z+u_z^4
\nonumber\\&
      +u_z^{-2}\chi^{E_7}_{\bm{7371}})q^4
+(-2-u_z^3 -\chi^{E_7}_{\bm{133}})\chi^J_1q^{\frac{9}{2}}
+(2u_z^{-1}\underline{+u_z^{-1}\chi^F_4}+u_z^2+u_z^{-1}\chi^J_2
\nonumber\\&
    -u_z^{-1}\chi^{E_7}_{\bm{133}}
    +u_z^{-1}\chi^J_2\chi^{E_7}_{\bm{133}}
    -u_z^{-1}\chi^{E_7}_{\bm{7371}}
    -u_z^{-1}\chi^{E_7}_{\bm{8645}})q^5
+\cdots,\\
%%%%%%%%%%%%%%%%%%%%%%%%%%%%%%%%%%%%%%%%%
{\cal I}_{E_8[1]}^{\rm AdS}
&=1
+u_z^{-1}\chi^{E_8}_{\bm{248}}q^2
+(-1-\chi^{E_8}_{\bm{248}})q^3
+(u_z^{-1}+u_z^{-1}\chi^{E_8}_{\bm{248}})\chi^J_1q^{\frac{7}{2}}
+(u_z+u_z^2\chi^{E_8}_{\bm{27000}})q^4
\nonumber\\&
+(-2-\chi^{E_8}_{\bm{248}})\chi^J_1q^{\frac{9}{2}}
+(2u_z^{-1}\underline{+u_z^{-1}\chi^F_4}+u_z^{-1}\chi^J_2
   -u_z^{-1}\chi^{E_8}_{\bm{248}}+u_z^{-1}\chi^J_2\chi^{E_8}_{\bm{248}}
\nonumber\\&
   -u_z^{-1}\chi^{E_8}_{\bm{27000}}
   -u_z^{-1}\chi^{E_8}_{\bm{30380}})q^5
+\cdots.
\end{align}
\end{subequations}

Because we took only the leading corrections into account
these results have errors, and it is important to determine
their orders.
Even without comparing these with the known results,
we can find terms that cannot be correct.
There are two types of such impossible terms.
The first type includes terms depending on the $SU(2)_F$ fugacity $u$.
Because $SU(2)_F$ symmetry decouples in the $N=1$ case
the index must be $u$-independent.
Indeed, many $u$-dependent terms appearing in ${\cal I}^{\rm KK}_G$
shown in Appendix \ref{largen.app} are drastically canceled
by the single-wrapping contributions.
For example,
the $u$-dependent term $q^2u_z^{-1}\chi_2^F$ appearing in ${\cal I}^{\rm KK}_G$ for all $G$
is canceled by
the leading term of the finite $N$ correction (\ref{theleadoing}) with $N=1$.
Even so, there still exist
terms with non-trivial $SU(2)_F$ characters $\chi^F_{n>0}$,
which must be canceled by higher order corrections.
The second type of impossible terms are terms diverging in the Coulomb branch limit
\begin{align}
q\rightarrow 0
\quad\mbox{with}\quad
qu_x^{-2},\quad
qu_y^{-2},\quad
qu_z,\quad
y
\quad\mbox{fixed}.
\end{align}
In this limit the $q$-expansion of ${\cal I}_{(1,0,0)}$
includes diverging terms
originating from the diverging factor $(qu_x)^{-1}$ in
$i({\cal B}^{{\rm D3}(X=0)}_{\frac{m}{2},r})$. (See (\ref{ibcd3}).)
Such diverging terms appearing in  (\ref{sciresults})
must also be canceled by higher order corrections.
We put underlines on the impossible terms in (\ref{sciresults}).

The lowest order impossible terms appearing in (\ref{sciresults}) are
\begin{align}
u_z^3(u_zq)^{4\Delta_G-2}\quad\mbox{for $G=H_0,H_1,H_2$},\quad
u_z^{-1}\chi^F_4q^5\quad\mbox{for $G=D_4,E_6,E_7,E_8$}.
\label{inconsi}
\end{align}
(In addition, we have another impossible term $u_z^2\chi_2^Fq^5$ in the $D_4$ case.)
If we assume these are the lowest order error terms, we can read off the tachyonic shifts
for the next-to-leading corrections
\begin{align}
\delta_{H_n}=3\Delta_G-2,\quad
\delta_{D_4}=\delta_{E_n}=3
\label{scitshift}
\end{align}
by comparing (\ref{errororder}) and (\ref{inconsi}).
Indeed, by comparing (\ref{sciresults})
and the known results
in the literature, we can confirm this is the case
except for $G=E_8$,
for which the superconformal index is knot known.
See Table \ref{error.tbl}.
\begin{table}[htb]
\caption{The differences between results in (\ref{sciresults}) calculated by the approximate formula
(\ref{theformula}) and the previously known results in the references are shown.}\label{error.tbl}
\centering
\begin{tabular}{llc}
\hline
\hline
Theory & ${\cal I}_{G[1]}-{\cal I}_{G[1]}^{AdS}$ & Refs. \\
\hline
$H_0[1]=(A_1,A_2)$ & $-u_z^{\frac{29}{5}}q^{\frac{14}{5}}+\cdots$ & \cite{Maruyoshi:2016tqk} \\
$H_1[1]=(A_1,A_3)$ & $-u_z^{\frac{19}{3}}q^{\frac{10}{3}}+\cdots$ & \cite{Maruyoshi:2016aim} \\
$H_2[1]=(A_1,D_4)$ & $-(u_z^4+u_z^7)q^{4}+\cdots$ & \cite{Agarwal:2016pjo} \\
$D_4[1]$ & $-(u_z^{-1}+u_z^{-1}\chi^F_4+u_z^2\chi^F_2+u_z^5)q^5+\cdots$ \\
$E_6[1]$ & $-(u_z^{-1}+u_z^{-1}\chi^F_4)q^5+\cdots$ & \cite{Gadde:2010te} \\
$E_7[1]$ & $-(u_z^{-1}+u_z^{-1}\chi^F_4)q^5+\cdots$ & \cite{Agarwal:2018ejn} \\
$E_8[1]$ & $?$ & unknown \\
\hline
\end{tabular}
\end{table}

It is also instructive to consider some limits
simplifying the structure of the index \cite{Gadde:2011uv}.
The Hall-Littlewood index is defined by taking the limit
\begin{align}
q\rightarrow 0
\quad\mbox{with}\quad
qu_x,\quad
qu_y,\quad
\olq\equiv qu_z^{-\frac{1}{2}},\quad
y\quad
\mbox{fixed}.
\end{align}
In this limit the Kaluza-Klein contribution becomes
\begin{align}
{\cal I}^{\rm KK}_{G}|_{\rm HL}
=\Pexp
\left(\sum_{k=2}^\infty \olq^k(\chi^F_k+\chi^F_{k-2}\chi^G_\theta)\right)
=1+\olq^2(\chi^F_2+\chi^G_\theta)+\olq^3(\chi_3^F+\chi^F_1\chi^G_\theta)+\cdots.
\label{largenhw}
\end{align}
By using
$i^{{\rm D3}(X=0)}_G|_{\rm HL}=\frac{1}{\olq u}\frac{1}{1-\olq u^{-1}}$ and
$\chi^{\wh G}(\olq u^{-1})=1+\olq u^{-1}\chi^G_\theta+\cdots$,
we obtain
\begin{align}
{\cal I}_{(1,0,0)}+{\cal I}_{(0,1,0)}|_{\rm HL}
=-\olq^2\chi^F_2-\olq^3(\chi^F_3+\chi^F_1\chi^G_\theta)+\cdots.
\label{d3n1}
\end{align}
We can easily see that
the $u$-dependent terms shown in
(\ref{largenhw}) and
(\ref{d3n1})
at the order $\olq^2$ and $\olq^3$ cancel.
This cancellation occurs also at $\olq^4$ and $\olq^5$,
and we find the first $u$-dependence at $\olq^6$.
We obtain the following result.
\begin{align}
{\cal I}_{G[1]}^{AdS}|_{\rm HL}
=1+\olq^2\chi^G_\theta+\olq^4\chi^G_{2\theta}\underline{+\olq^6(\mbox{$u$-dep.})}+\cdots
\label{rank1hl}
\end{align}
We can also calculate the Schur index
in a similar way.
The Schur index is defined from the
superconformal index by setting
$y=q^{\frac{1}{2}}u_z^{-1}$, and
is a function of
$\olq=qu_z^{-\frac{1}{2}}$, $u$, and $G$ fugacities.
We obtain
\begin{align}
{\cal I}^{\rm AdS}_{G[1]}|_{\rm Sch}=1+\olq^2\chi^G_\theta+\olq^4(1+\chi^G_\theta+\chi^G_{2\theta})
\underline{+\olq^6(\mbox{$u$-dep.})}+\cdots.
\label{rank1schur}
\end{align}
Unlike the superconformal index these limits of the index do not acquire
the contribution from
the $Z=0$ cycle, and the next-to-leading corrections should be of order $q^{2N+\delta_G}$.
(\ref{rank1hl}) and (\ref{rank1schur})
suggest that the higher order corrections start at $\olq^6$.
We can confirm this by comparing
(\ref{rank1hl}) and 
(\ref{rank1schur}) with known results \cite{Benvenuti:2010pq,Keller:2011ek,Keller:2012da,Buican:2015ina,Cordova:2015nma,Song:2015wta}.
This means that the tachyonic shift for the Hall-Littlewood index and the Schur index is
\begin{align}
\delta_G=4
\label{hlschtshift}
\end{align}
for all $G$.

%%%%%%%%%%%%%%%%%%%%%%%%%%%%%%%%%%%%%%%%
\section{Higher rank theories}\label{higherrank.sec}
An advantage of the method using AdS/CFT is that
we can deal with higher rank theories in
the same way as the rank $1$ theories.
In this section we show the results for $N=2$ and $N=3$. 
Many of these have been calculated in the literature
\cite{Hanany:2012dm,Gu:2019dan,Beem:2019snk,Beem:2020pry},
and our results are consistent with them.

We determine the order of error terms,
which are indicated by underlines in the following results,
by using the tachyonic shifts (\ref{scitshift})
for the superconformal index and (\ref{hlschtshift})
for the Hall-Littlewood index and Schur index.
Namely, the order of the error terms for rank $N$ are
\begin{align}
{\cal O}(q^{\Delta_G(N+3)-2})\quad(G=H_n),\quad
{\cal O}(q^{2N+3}),\quad(G=D_4,E_n).
\label{scierrororders}
\end{align}
for the superconformal index and
\begin{align}
{\cal O}(\olq^{2N+4})
\label{schurerrorders}
\end{align}
for the Hall-Littlewood and Schur indices

\subsection{Superconformal index}
\begin{subequations}
\begin{align}
{\cal I}_{H_0[2]}^{\rm AdS}
&\eqo 1
+q^{\frac{6}{5}}
-2q^{\frac{17}{10}}
+3q^2
+3q^{\frac{11}{5}}
+2q^{\frac{12}{5}}
-2q^{\frac{27}{10}}
-4q^{\frac{29}{10}}
-4q^3
-q^{\frac{16}{5}}
+5q^{\frac{17}{5}}
\nonumber\\&
+8q^{\frac{7}{2}}
+2q^{\frac{18}{5}}
+4q^{\frac{37}{10}}
-6q^{\frac{39}{10}}
\underline{+5q^4}
+\cdots,\\
%%%%%%%%%%%%%%%%%%%%%%%%%%%%%%%%%%%%%%%
{\cal I}_{H_1[2]}^{\rm AdS}
&\eqo 1
+q^{\frac{4}{3}}
-2q^{\frac{11}{6}}
+6q^2
+3q^{\frac{7}{3}}
+2q^{\frac{8}{3}}
-2q^{\frac{17}{6}}
-q^3
-4q^{\frac{19}{6}}
+2q^{\frac{10}{3}}
+14q^{\frac{7}{2}}
\nonumber\\&
+5q^{\frac{11}{3}}
-2q^{\frac{23}{6}}
+15q^4
-6q^{\frac{25}{6}}
+q^{\frac{13}{3}}
-6q^{\frac{9}{2}}
\underline{+7q^{\frac{14}{3}}}
+\cdots,\\
%%%%%%%%%%%%%%%%%%%%%%%%%%%%%%%%%%%%%%%
{\cal I}_{H_2[2]}^{\rm AdS}
&\eqo 1
+q^{\frac{3}{2}}
+9q^2
+3q^{\frac{5}{2}}
+4q^3
+27q^{\frac{7}{2}}
+41q^4
+17q^{\frac{9}{2}}
+81q^5
\underline{+183q^{\frac{11}{2}}}
+\cdots,\\
%%%%%%%%%%%%%%%%%%%%%%%%%%%%%%%%%%%%%%%
{\cal I}_{D_4[2]}^{\rm AdS}
&\eqo 1
+32q^2
-2q^{\frac{5}{2}}
+27q^3
+62q^{\frac{7}{2}}
+467q^4
-6q^{\frac{9}{2}}
+632q^5
+1924q^{\frac{11}{2}}
\nonumber\\&
+3702q^6
+2326q^{\frac{13}{2}}
\underline{+8420q^7}
+\cdots,\\
%%%%%%%%%%%%%%%%%%%%%%%%%%%%%%%%%%%%%%%
{\cal I}_{E_6[2]}^{\rm AdS}
&\eqo 1
+81q^2
+75q^3
+162q^{\frac{7}{2}}
+3166q^4
+148q^{\frac{9}{2}}
+4863q^5
+12812q^{\frac{11}{2}}
\nonumber\\&
+78247q^6
+21552q^{\frac{13}{2}}
\underline{+158937q^7}
+\cdots,\\
%%%%%%%%%%%%%%%%%%%%%%%%%%%%%%%%%%%%%%%
{\cal I}_{E_7[2]}^{\rm AdS}
&\eqo 1
+136q^2
+129q^3
+274q^{\frac{7}{2}}
+9049q^4
+258q^{\frac{9}{2}}
+14616q^5
+36722q^{\frac{11}{2}}
\nonumber\\&
+389749q^6
+63772q^{\frac{13}{2}}
\underline{+825721q^7}
+\cdots,\\
%%%%%%%%%%%%%%%%%%%%%%%%%%%%%%%%%%%%%%%
{\cal I}_{E_8[2]}^{\rm AdS}
&\eqo 1
+251q^2
+244q^3
+504q^{\frac{7}{2}}
+31128q^4
+490q^{\frac{9}{2}}
+53756q^5
+125504q^{\frac{11}{2}}
\nonumber\\&
+2539245q^6
+229488q^{\frac{13}{2}}
\underline{+5896389q^7}+\cdots
\end{align}
\end{subequations}

\begin{subequations}
\begin{align}
{\cal I}_{H_0[3]}^{\rm AdS}
&\eqo 1
+q^{\frac{6}{5}}
-2q^{\frac{17}{10}}
+3q^2
+3q^{\frac{11}{5}}
+2q^{\frac{12}{5}}
-2q^{\frac{27}{10}}
-4q^{\frac{29}{10}}
+2q^{\frac{16}{5}}
+7q^{\frac{17}{5}}
+8q^{\frac{7}{2}}
\nonumber\\&
+3q^{\frac{18}{5}}
-2q^{\frac{37}{10}}
-10q^{\frac{39}{10}}
+q^4
-8q^{\frac{41}{10}}
+q^{\frac{21}{5}}
+11q^{\frac{22}{5}}
+2q^{\frac{9}{2}}
+15q^{\frac{23}{5}}
+10q^{\frac{47}{10}}
\nonumber\\&
+4q^{\frac{24}{5}}
-2q^{\frac{49}{10}}
+10q^5
-22q^{\frac{51}{10}}
\underline{-21q^{\frac{26}{5}}}
+\cdots,\\
%%%%%%%%%%%%%%%%%%%%%%%%%%%%%%%%%%%%%%%
{\cal I}_{H_1[3]}^{\rm AdS}
&\eqo 1
+q^{\frac{4}{3}}
-2q^{\frac{11}{6}}
+6q^2
+3q^{\frac{7}{3}}
+2q^{\frac{8}{3}}
-2q^{\frac{17}{6}}
+3q^3
-4q^{\frac{19}{6}}
+5q^{\frac{10}{3}}
+14q^{\frac{7}{2}}
\nonumber\\&
+7q^{\frac{11}{3}}
-8q^{\frac{23}{6}}
+22q^4
-10q^{\frac{25}{6}}
+13q^{\frac{13}{3}}
+17q^{\frac{14}{3}}
+4q^{\frac{29}{6}}
+43q^5
-14q^{\frac{31}{6}}
\nonumber\\&
-15q^{\frac{16}{3}}
+60q^{\frac{11}{2}}
+7q^{\frac{17}{3}}
+32q^{\frac{35}{6}}
\underline{+81q^6}
+\cdots,\\
%%%%%%%%%%%%%%%%%%%%%%%%%%%%%%%%%%%%%%%
{\cal I}_{H_2[3]}^{\rm AdS}
&\eqo 1
+q^{\frac{3}{2}}
+9q^2
+3q^{\frac{5}{2}}
+8q^3
+30q^{\frac{7}{2}}
+58q^4
+44q^{\frac{9}{2}}
+111q^5
+259q^{\frac{11}{2}}
\nonumber\\&
+374q^6
+462q^{\frac{13}{2}}
\underline{+1000q^7}
+\cdots,\\
%%%%%%%%%%%%%%%%%%%%%%%%%%%%%%%%%%%%%%%
{\cal I}_{D_4[3]}^{\rm AdS}
&=1
+32q^2
-2q^{\frac{5}{2}}
+31q^3
+62q^{\frac{7}{2}}
+551q^4
-4q^{\frac{9}{2}}
+998q^5
+1976q^{\frac{11}{2}}
\nonumber\\&
+6661q^6
+2862q^{\frac{13}{2}}
+17537q^7
+35482q^{\frac{15}{2}}
+64679q^8
+84630q^{\frac{17}{2}}
\nonumber\\&\quad
\underline{+220412q^9}
+\cdots,\\
%%%%%%%%%%%%%%%%%%%%%%%%%%%%%%%%%%%%%%%
{\cal I}_{E_6[3]}^{\rm AdS}
&\eqo 1
+81q^2
+79 q^3
+162q^{\frac{7}{2}}
+3397q^4
+156q^{\frac{9}{2}}
+6408q^5
+13274q^{\frac{11}{2}}
\nonumber\\&
+99165q^6
+25290q^{\frac{13}{2}}
+273109q^7
+570728q^{\frac{15}{2}}
+2283657q^8
\nonumber\\&
+1549838q^{\frac{17}{2}}
\underline{+8097884q^9}
+\cdots,\\
%%%%%%%%%%%%%%%%%%%%%%%%%%%%%%%%%%%%%%%
{\cal I}_{E_7[3]}^{\rm AdS}
&\eqo 1
+136q^2
+133q^3
+274q^{\frac{7}{2}}
+9445q^4
+266q^{\frac{9}{2}}
+18101q^5
+37520q^{\frac{11}{2}}
\nonumber\\&
+450243q^6
+72362q^{\frac{13}{2}}
+1271046q^7
+2647564q^{\frac{15}{2}}
+16686266q^8
\nonumber\\&
+7415174q^{\frac{17}{2}}
\underline{+61224202q^9}
+\cdots,\\
%%%%%%%%%%%%%%%%%%%%%%%%%%%%%%%%%%%%%%%
{\cal I}_{E_8[3]}^{\rm AdS}
&\eqo 1
+251q^2
+248q^3
+504q^{\frac{7}{2}}
+31869q^4
+498q^{\frac{9}{2}}
+62258q^5
+126992q^{\frac{11}{2}}
\nonumber\\&
+2747126q^6
+249498q^{\frac{13}{2}}
+7961389q^7
+16282232q^{\frac{15}{2}}
+181906110q^8
\nonumber\\&
+47084068q^{\frac{17}{2}}
\underline{+691172658q^9}
+\cdots.
\end{align}
\end{subequations}

%%%%%%%%%%%%%%%%%%%%%%%%%%%%%%%%%%%%%%%%%
\subsection{Hall-Littlewood index}

\begin{subequations}
\begin{align}
{\cal I}^{\rm AdS}_{H_0[2]}|_{\rm HL}
&\eqo 1 + 3\olq^2 + 5\olq^4 + 7\olq^6 \underline{+ 30\olq^8}
+\cdots,\\
%%%%%%%%%%%%%%%%%%%%%%%%%%%%%%%%%%%%%%%
{\cal I}^{\rm AdS}_{H_1[2]}|_{\rm HL}
&\eqo 1+6\olq^2 + 6\olq^3 + 20\olq^4 + 28\olq^5 + 65\olq^6 + 80\olq^7
\underline{+ 242\olq^8}
+\cdots,\\
%%%%%%%%%%%%%%%%%%%%%%%%%%%%%%%%%%%%%%%
{\cal I}^{\rm AdS}_{H_2[2]}|_{\rm HL}
&\eqo 1+11\olq^2 + 16\olq^3 + 65\olq^4 + 142\olq^5 + 355\olq^6 + 700\olq^7
\underline{+ 1779\olq^8}
+\cdots,\\
%%%%%%%%%%%%%%%%%%%%%%%%%%%%%%%%%%%%%%%
{\cal I}^{\rm AdS}_{D_4[2]}|_{\rm HL}
&\eqo 1+31\olq^2 + 56\olq^3 + 495\olq^4 + 1468\olq^5 + 6269\olq^6 + 19680\olq^7
\underline{+66611\olq^8}
+\cdots,\\
%%%%%%%%%%%%%%%%%%%%%%%%%%%%%%%%%%%%%%%
{\cal I}^{\rm AdS}_{E_6[2]}|_{\rm HL}
&\eqo 1 + 81\olq^2 + 156\olq^3 + 3320\olq^4 + 11178\olq^5 + 98440\olq^6 +
 401280\olq^7
\nonumber\\&
\underline{+ 2356455\olq^8}
+\cdots,\\
%%%%%%%%%%%%%%%%%%%%%%%%%%%%%%%%%%%%%%%
{\cal I}^{\rm AdS}_{E_7[2]}|_{\rm HL}
&\eqo 1+136\olq^2 + 266\olq^3 + 9315\olq^4 + 32830\olq^5 + 449050\olq^6
+ 2026080\olq^7
\nonumber\\&
\underline{+ 17206093\olq^8}
+\cdots,\\
%%%%%%%%%%%%%%%%%%%%%%%%%%%%%%%%%%%%%%%
{\cal I}^{\rm AdS}_{E_8[2]}|_{\rm HL}
&\eqo 1+251\olq^2 + 496\olq^3 + 31625\olq^4 + 116248\olq^5 + 2747875\olq^6
 + 13624000\olq^7
\nonumber\\&
\underline{+ 187007628\olq^8}
+\cdots.
\end{align}
\end{subequations}

\begin{subequations}
\begin{align}
{\cal I}^{\rm AdS}_{H_0[3]}|_{\rm HL}
&\eqo 1+3\olq^2 + 4\olq^3 + 6\olq^4 + 10\olq^5 + 17\olq^6 + 18\olq^7 + 31\olq^8 + 38\olq^9
\underline{+76\olq^{10}}
+\cdots,\\
%%%%%%%%%%%%%%%%%%%%%%%%%%%%%%%%%%%%%%%%%
{\cal I}^{\rm AdS}_{H_1[3]}|_{\rm HL}
&\eqo 1+6\olq^2 + 10\olq^3 + 30\olq^4 + 58\olq^5 + 147\olq^6 + 258\olq^7 + 548\olq^8 +
 952\olq^9
\nonumber\\&
\underline{+ 1876\olq^{10}}
+\cdots,\\
%%%%%%%%%%%%%%%%%%%%%%%%%%%%%%%%%%%%%%%%%
{\cal I}^{\rm AdS}_{H_2[3]}|_{\rm HL}
&\eqo 1+11\olq^2 + 20\olq^3 + 90\olq^4 + 218\olq^5 + 698\olq^6 + 1618\olq^7 + 4300\olq^8 +
 9588\olq^9
\nonumber\\&
 \underline{+ 22634\olq^{10}}
+\cdots,\\
%%%%%%%%%%%%%%%%%%%%%%%%%%%%%%%%%%%%%%%%%
{\cal I}^{\rm AdS}_{D_4[3]}|_{\rm HL}
&\eqo 1+31\olq^2 + 60\olq^3 + 580\olq^4 + 1858\olq^5 + 9457\olq^6 + 33066\olq^7 +
 131755\olq^8
\nonumber\\&
+ 444502\olq^9
\underline{+ 1543882\olq^{10}}
+\cdots,\\
%%%%%%%%%%%%%%%%%%%%%%%%%%%%%%%%%%%%%%%%%
{\cal I}^{\rm AdS}_{E_6[3]}|_{\rm HL}
&\eqo 1+81\olq^2 + 160\olq^3 + 3555\olq^4 + 12958\olq^5 + 121447\olq^6 + 556958\olq^7 +
 3563694\olq^8
\nonumber\\&
 + 17126502\olq^9
 \underline{+ 90513091\olq^{10}}
+\cdots,\\
%%%%%%%%%%%%%%%%%%%%%%%%%%%%%%%%%%%%%%%%%
{\cal I}^{\rm AdS}_{E_7[3]}|_{\rm HL}
&\eqo 1+136\olq^2 + 270\olq^3 + 9715\olq^4 + 36718\olq^5 + 514230\olq^6 +
 2592258\olq^7
\nonumber\\&
 + 22872825\olq^8 + 128145440\olq^9
 \underline{+ 885685093\olq^{10}}
+\cdots,\\
%%%%%%%%%%%%%%%%%%%%%%%%%%%%%%%%%%%%%%%%%
{\cal I}^{\rm AdS}_{E_8[3]}|_{\rm HL}
&\eqo 1+251\olq^2 + 500\olq^3 + 32370\olq^4 + 125498\olq^5 + 2966497\olq^6 +
 16098498\olq^7
\nonumber\\&
 + 221148375\olq^8 + 1420026502\olq^9
\underline{+ 14229178180\olq^{10}}
+\cdots.
\end{align}
\end{subequations}

%%%%%%%%%%%%%%%%%%%%%%%%%%%%%%%%%%%%%%%%%%%%%%%%%%%%%%%%%
\subsection{Schur index}

\begin{subequations}
\begin{align}
{\cal I}_{H_0[2]}^{\rm AdS}|_{\rm Sch}
&\eqo 1+3\olq^2+9\olq^4+2\olq^5+22\olq^6+6\olq^7\underline{+62\olq^8}
+\cdots,\\
%%%%%%%%%%%%%%%%%%%%%%%%%%%%%%%%%%%%%%%
{\cal I}_{H_1[2]}^{\rm AdS}|_{\rm Sch}
&\eqo 1+6\olq^2+6\olq^3+27\olq^4+36\olq^5+113\olq^6+162\olq^7\underline{+471\olq^8}
+\cdots,\\
%%%%%%%%%%%%%%%%%%%%%%%%%%%%%%%%%%%%%%%
{\cal I}_{H_2[2]}^{\rm AdS}|_{\rm Sch}
&\eqo 1+11\olq^2+16\olq^3+77\olq^4+160\olq^5+498\olq^6+1056\olq^7\underline{+2950\olq^8}
+\cdots,\\
%%%%%%%%%%%%%%%%%%%%%%%%%%%%%%%%%%%%%%%
{\cal I}_{D_4[2]}^{\rm AdS}|_{\rm Sch}
&\eqo 1+31\olq^2+56\olq^3+527\olq^4+1526\olq^5+7292\olq^6+23002\olq^7\underline{+86239\olq^8}
+\cdots,\\
%%%%%%%%%%%%%%%%%%%%%%%%%%%%%%%%%%%%%%%
{\cal I}_{E_6[2]}^{\rm AdS}|_{\rm Sch}
&\eqo 1+81\olq^2+156\olq^3+3402\olq^4+11336\olq^5+105163\olq^6+425412\olq^7
\nonumber\\&
\underline{+2656809\olq^8}
+\cdots,\\
%%%%%%%%%%%%%%%%%%%%%%%%%%%%%%%%%%%%%%%
{\cal I}_{E_7[2]}^{\rm AdS}|_{\rm Sch}
&\eqo 1+136\olq^2+266\olq^3+9452\olq^4+33098\olq^5+467818\olq^6+2095624\olq^7
\nonumber\\&
\underline{+18564678\olq^8}
+\cdots,\\
%%%%%%%%%%%%%%%%%%%%%%%%%%%%%%%%%%%%%%%
{\cal I}_{E_8[2]}^{\rm AdS}|_{\rm Sch}
&\eqo 1+251\olq^2+496\olq^3+31877\olq^4+116746\olq^5+2811378\olq^6+13865742\olq^7
\nonumber\\&
\underline{+195272132\olq^8}
+\cdots.
\end{align}
\end{subequations}

\begin{subequations}
\begin{align}
{\cal I}_{H_0[3]}^{\rm AdS}|_{\rm Sch}
&\eqo 1+3\olq^2+4\olq^3+10\olq^4+16\olq^5+36\olq^6+56\olq^7+110\olq^8+176\olq^9
\nonumber\\&
\underline{+327\olq^{10}}
+\cdots,\\
%%%%%%%%%%%%%%%%%%%%%%%%%%%%%%%%%%%%%%%
{\cal I}_{H_1[3]}^{\rm AdS}|_{\rm Sch}
&\eqo 1+6\olq^2+10\olq^3+37\olq^4+70\olq^5+208\olq^6+410\olq^7+1008\olq^8+2000\olq^9
\nonumber\\&
\underline{+4501\olq^{10}}
+\cdots,\\
%%%%%%%%%%%%%%%%%%%%%%%%%%%%%%%%%%%%%%%
{\cal I}_{H_2[3]}^{\rm AdS}|_{\rm Sch}
&\eqo 1+11\olq^2+20\olq^3+102\olq^4+240\olq^5+869\olq^6+2120\olq^7+6276\olq^8
\nonumber\\&
+15220\olq^9+\underline{+40356\olq^{10}}
+\cdots,\\
%%%%%%%%%%%%%%%%%%%%%%%%%%%%%%%%%%%%%%%
{\cal I}_{D_4[3]}^{\rm AdS}|_{\rm Sch}
&\eqo 1+31\olq^2+60\olq^3+612\olq^4+1920\olq^5+10568\olq^6+36968\olq^7+157850\olq^8
\nonumber\\&
+548848\olq^9+\underline{+2039418\olq^{10}}
+\cdots,\\
%%%%%%%%%%%%%%%%%%%%%%%%%%%%%%%%%%%%%%%
{\cal I}_{E_6[3]}^{\rm AdS}|_{\rm Sch}
&\eqo 1+81\olq^2+160\olq^3+3637\olq^4+13120\olq^5+128408\olq^6+583360\olq^7
\nonumber\\&
+3908179\olq^8+18828800\olq^9+\underline{+103829612\olq^{10}}
+\cdots,\\
%%%%%%%%%%%%%%%%%%%%%%%%%%%%%%%%%%%%%%%
{\cal I}_{E_7[3]}^{\rm AdS}|_{\rm Sch}
&\eqo 1+136\olq^2+270\olq^3+9852\olq^4+36990\olq^5+533401\olq^6+2666510\olq^7
\nonumber\\&
+24354958\olq^8+136003400\olq^9+\underline{+972032920\olq^{10}}
+\cdots,\\
%%%%%%%%%%%%%%%%%%%%%%%%%%%%%%%%%%%%%%%
{\cal I}_{E_8[3]}^{\rm AdS}|_{\rm Sch}
&\eqo 1+251\olq^2+500\olq^3+32622\olq^4+126000\olq^5+3030748\olq^6+16351000\olq^7
\nonumber\\&
+229826870\olq^8+1468558000\olq^9+\underline{+15077246917\olq^{10}}
+\cdots.
\end{align}
\end{subequations}

%%%%%%%%%%%%%%%%%%%%%%%%%%%%%%%%%%%%%%%%%%%%%%%%%%%%%%%%%%%%%%%%%%%%%%%%%%%%%%%%%%%
\section{Conclusions and discussion}\label{disc.sec}
In this paper we calculated the superconformal indices of the $\mathcal N=2$ theories realized by D3-7-brane systems.
We first confirmed the method works well in the $D_4$ case,
for which we can perturbatively calculate the index both on the gauge theory side and on the AdS side,
and then we applied the same method to more interesting cases with $G\neq D_4$.
The results are consistent with known results in the literature,
and some of them are new, and give predictions.

In this paper we focused only on ${\cal I}_{(1,0,0)}$ and ${\cal I}_{(0,1,0)}$,
the contributions of single-wrapping D3-branes around two cycles $X=0$ and $Y=0$, for simplicity.
To improve our results we need to include higher order contributions.
The orders of the next-to-leading corrections are given in (\ref{scierrororders})
for the superconformal index and
in (\ref{schurerrorders}) for the Schur and Hall-Littlewood indices.
For the latter, the $Z=0$ cycle does not contribute and
multiple-wrapping D3-branes with $n_x+n_y\geq 2$ give the higher order corrections.
As was studied in \cite{Arai:2020qaj} and \cite{Imamura:2021ytr}, in order to calculate multiple-wrapping contributions,
we need to choose very carefully the integration contours in the holonomy integrals.
In addition, we need to take account of the current algebra localized on the
intersection of the D3-brane and the 7-brane.
In the case of the superconformal index, the $Z=0$ cycle also contribute.
If $n_z$ D3-branes are wrapped around the cycle, the theory realized on the worldvolume is
$G[n_z]$, and perturbative treatment is not possible except for the $D_4$ case.
Although direct calculation of such a contribution is
difficult,
it may be possible to extract some information about them from the error obtained in
our analysis.
For example, the tachyonic shifts $\delta_{H_n}=3\Delta_G-2$ in (\ref{scitshift})
should be somehow interpreted in $H_n[1]$ on the wrapped D3.

Recently,
an expansion similar to (\ref{expansion}) was proposed in \cite{Gaiotto:2021xce} for
Lagrangian gauge theories based on the direct analysis in the gauge theories.
Unlike the multiple expansion (\ref{expansion}) the expansion
in \cite{Gaiotto:2021xce} is a simple expansion.
It is interesting question whether such a simple expansion
exists for non-Lagrangian theories like ones we studied in this paper.

Another important direction is to consider more complicated background geometries.
Although only limited class of Argyres-Dougras theories are realized by the D3-7-brane systems,
more general Argyres-Douglas theories can be realized
\cite{Xie:2012hs}
as class S theories \cite{Gaiotto:2009we,Gaiotto:2009hg},
and some supergravity solutions have been proposed \cite{Gaiotto:2009gz,Bah:2021mzw,Bah:2021hei}.
It would be interesting to study to what extent the method can be applied
in such more complicated backgrounds.

%%%%%%%%%%%%%%%%%%%%%%%%%%%%%%%%%%%%%%%%%%%%%%%%%%%%%%%%%%%%%%%%%%%%%%%%%%%%%%%%%%%
\section*{Acknowledgments}
We would like to thank Reona Arai, Shota Fujiwara, Tatsuya Mori,
and Daisuke Yokoyama for valuable discussions and comments.
The work of Y.~I. was
partially supported by Grand-in-Aid for Scientific Research (C) (No.21K03569),
Ministry of Education, Culture, Sports, Science and Technology, Japan.

\appendix
%%%%%%%%%%%%%%%%%%%%%%%%%%%%%%%%%%%%%%%%%%%%%%%%%%%%%%%%%%
\section{Gravity multiplet}\label{grav.app}

Kaluza-Klein modes of gravity multiplet in $AdS_5\times S^5$ belong to
the representations ${\cal B}^{\frac{1}{2},\frac{1}{2}}_{[0,n,0](0,0)}$ ($n=1,2,3,\ldots$).
We use notation in \cite{Dolan:2002zh} for superconformal representations.
Each of them are constructed by acting ${\cal N}=4$ supercharges
on the lowest energy states belonging to the $SU(4)_R$ representation
$[0,n,0]$.
The $SU(2)_R\times SU(2)_F\times U(1)_{R_Z}$ decomposition of this representation is
\begin{align}
[0,n,0]\rightarrow\bigoplus_{m+k+l=n}(\tfrac{m}{2},\tfrac{m}{2})_{k-l}
\label{dec22i}
\end{align}
where the direct sum is taken over partitions of $n$ into three non-negative
integers $m$, $k$, and $l$.
The ${\cal N}=2$ representations are obtained by acting supercharges on the representations
appearing in (\ref{dec22i}).
Although there may be other superconformal representations whose primaries
do not belong to (\ref{dec22i}),
such representations do not contribute to the index and
we can neglect them.
Furthermore, only representations with $l=0$ and $l=1$ contribute to the index.
The supersymmetry completion of the relevant representations are
\begin{align}
(\tfrac{m}{2},{\tfrac{m}{2}})_k\rightarrow
{\cal B}_{\frac{m}{2},k(0,0)}\otimes[\tfrac{m}{2}]_F,\quad
(\tfrac{m}{2},{\tfrac{m}{2}})_{k-1}\rightarrow
{\cal C}_{\frac{m}{2},k-1(0,0)}\otimes[\tfrac{m}{2}]_F.
\label{bcseries}
\end{align}
Both $k$ and $m$ run over non-negative integers.
Representations with $k=0$ and $k=1$ have special structure and
they are denoted in \cite{Dolan:2002zh}
as follows:
\begin{align}
{\cal B}_{\frac{m}{2},0(0,0)}
=\wh{\cal B}_{\frac{m}{2}},\quad
{\cal B}_{\frac{m}{2},1(0,0)}
=\wh{\cal D}_{\frac{m}{2}(0,0)},\quad
{\cal C}_{\frac{m}{2},-1(0,0)}
=\ol{\cal D}_{\frac{m}{2}(0,0)},\quad
{\cal C}_{\frac{m}{2},0(0,0)}
=\wh{\cal C}_{\frac{m}{2}(0,0)}.
\end{align}
For the analysis of $G=H_n$
we need to extend the range of $k$ \cite{Aharony:1998xz}
to fractional values.
The unitarity requires $k=0$ or $k\geq1$.
This means $r=0$ or $r\geq1$ for ${\cal B}_{\frac{m}{2},r(0,0)}$
and $r=-1$ or $r\geq0$ for ${\cal C}_{\frac{m}{2},r(0,0)}$.
For each value of $r$ the contribution
of the representations in (\ref{bcseries})
with all allowed values of $m$
to the index are as follows:
\begin{align}
\sum_m i({\cal B}_{m,0(0,0)})\chi^F_m
&=\frac{1}{(\mbox{mom})}\left[\left\{\frac{1-qu_z}{(1-qu_x)(1-qu_y)}+qu_z\right\}-1\right],\nonumber\\
\sum_m i({\cal B}_{m,1(0,0)})\chi^F_m
&=\frac{1}{(\mbox{mom})}\left[(qu_z)(1-q^{\frac{1}{2}}y^{\pm1}u_z^{-1})
\left\{\frac{(1-qu_z)}{(1-qu_x)(1-qu_y)}+qu_z\right\}+q^3\right],\nonumber\\
\sum_m i({\cal B}_{m,r>1(0,0)})\chi^F_m
&=(qu_z)^r\frac{(1-q^{\frac{1}{2}}yu_z^{-1})(1-q^{\frac{1}{2}}y^{-1}u_z^{-1})}{\mbox{mom}}
\left\{\frac{1-qu_z}{(1-qu_x)(1-qu_y)}+qu_z\right\},
\nonumber\\
\sum_m i({\cal C}_{m,-1(0,0)})\chi^F_m
&=-\frac{1}{(\mbox{mom})}\frac{q^2u_xu_y(1-qu_z)}{(1-qu_x)(1-qu_y)},\nonumber\\
\sum_m i({\cal C}_{m,0(0,0)})\chi^F_m
&=-(qu_z)\frac{(1-q^{\frac{1}{2}}y^{\pm1}u_z^{-1})}{(\mbox{mom})}\frac{q^2u_xu_y(1-qu_z)}{(1-qu_x)(1-qu_y)},\nonumber\\
\sum_m i({\cal C}_{m,r>0(0,0)})\chi^F_m
&=-(qu_z)^{r+1}\frac{(1-q^{\frac{1}{2}}yu_z^{-1})
(1-q^{\frac{1}{2}}y^{-1}u_z^{-1})}{(\mbox{mom})}
\frac{q^2u_xu_y(1-qu_z)}{(1-qu_x)(1-qu_y)},
\end{align}
where $(\mbox{mom})$ is the momentum factor
\begin{align}
(\mbox{mom})=(1-q^{\frac{3}{2}}y)(1-q^{\frac{3}{2}}y^{-1}).
\end{align}

%%%%%%%%%%%%%%%%%%%%%%%%%%%%%%%%%%%%%%%%%%%%%%%%%%%%%
\section{Fluctuation modes on D3}\label{d3.app}
Let us consider a D3-brane wrapped on $X=0$.
The single-particle index for the fields living on the wrapped brane
in $S^5$ without deficit angle was calculated in \cite{Arai:2019xmp}
by using variable changes from the index of boundary ${\cal N}=4$ vector multiplet.
Explicit mode expansion is given in \cite{Imamura:2021ytr}.
Let us first review the derivation
and then we consider the effect of $7$-branes.

The existence of a D3-brane wrapped on $X=0$
respects only supercharges
with quantum numbers $H=R_X$,
and it breaks the superconformal algebra
$usp(2,2|4)$ to $psu(2|2)\times psu(2|2)$.
One of $psu(2|2)$ contains $SU(2)_J\times SU(2)_{\frac{R_Z-R_Y}{2}}$
and the other contains $SU(2)_{\ol J}\times SU(2)_{\frac{R_Z+R_Y}{2}}$
as the bosonic subgroups.
There are eight supercharges with positive conformal dimension:
$Q$ and $\ol Q$
belonging to the bi-fundamental representations of
$SU(2)_J\times SU(2)_{\frac{R_Z-R_Y}{2}}$
and $SU(2)_{\ol J}\times SU(2)_{\frac{R_Z+R_Y}{2}}$,
respectively.

We want to construct short multiplets of
excitations on the wrapped D3-brane.
We apply $Q$ and $\ol Q$ (and $SU(2)$ lowering operators)
as raising operators
on a ground state to form the whole multiplet.
Because all components of $Q$ and $\ol Q$ carry $R_X=+\frac{1}{2}$
the ground state must carry minimum $R_X$ in the multiplet.
We can use modes of $X^*$ and denote the corresponding ground state by $|X^*\rangle$.
There are different modes described by scalar $S^3$ harmonics
on the wrapped D3.
They are labeled by integer $\ell=0,1,2,\ldots$,
and belong to $[\frac{\ell}{2},\frac{\ell}{2}]$ of
$SU(2)_{\frac{R_Z-R_Y}{2}}\times SU(2)_{\frac{R_Z+R_Y}{2}}$.
We start from the
$SU(2)_{\frac{R_Z-R_Y}{2}}\times SU(2)_{\frac{R_Z+R_Y}{2}}$
highest weight state that carries
$\frac{R_Z-R_Y}{2}=\frac{R_Z+R_Y}{2}=\frac{\ell}{2}$,
or, equivalently, $R_Z=\ell$, $R_Y=0$.
By applying raising operators on this state we
obtain a $psu(2|2)\times psu(2|2)$ multiplet
for each $\ell$.

Due to the shortening conditions, we use only supercharges
that does not increase either $\frac{R_Z-R_Y}{2}$ or $\frac{R_Z+R_Y}{2}$.
Only two from $Q$ and two from $\ol Q$ satisfy this condition.
Furthermore, we are interested in BPS operators saturating
the bound (\ref{bpsbound}),
and hence we use only supercharges carrying $\{{\cal Q},{\cal Q}^\dagger\}=0$.
This condition excludes one of two components in $\ol Q$.
For the same reason we do not use the $SU(2)_{\frac{R_Z+R_Y}{2}}$ lowering operator.
As the result, we can use two components of $Q$, one component of $\ol Q$,
and
the $SU(2)_{\frac{R_Z-R_Y}{2}}$ lowering operator $R_-$
in the construction of the representation.
We show the quantum numbers of these operators and
the ground state in Table \ref{qqx}.
\begin{table}[htb]
\caption{Quantum numbers of the ground state and $Q$ and $\ol Q$ used as raising operators.}\label{qqx}
\centering
\begin{tabular}{cccccccc}
\hline
\hline
 & $H$ & $J$ & $\ol J$ & $R_X$ & $R_Y$ & $R_Z$ & $A$ \\
\hline
$Q$ & $+\frac{1}{2}$ & $\pm\frac{1}{2}$ & $0$ & $+\frac{1}{2}$ & $+\frac{1}{2}$ & $-\frac{1}{2}$ & $-1$ \\
$\ol Q$ & $+\frac{1}{2}$ & $0$ & $+\frac{1}{2}$ & $+\frac{1}{2}$ & $-\frac{1}{2}$ & $-\frac{1}{2}$ & $+1$ \\
$R_-$ & $0$ & $0$ & $0$ & $0$ & $+1$ & $-1$ & $0$ \\
$|X^*\rangle$ & $\ell-1$ & $0$ & $0$ & $-1$ & $0$ & $\ell$ & $0$ \\
\hline
\end{tabular}
\end{table}

A general state in the multiplet is schematically given as
\begin{align}
Q^p\ol Q^qR_-^m|X^*\rangle,\quad
p=0,1,2,\quad
q=0,1.
\label{qqrx}
\end{align}
The introduction of $7$-brane at $Z=0$ breaks supersymmetry $\ol Q$,
and as the result the states in (\ref{qqrx})
split into two types of representations:
${\cal B}^{{\rm D3}(X=0)}_{m,r}$
and ${\cal C}^{{\rm D3}(X=0)}_{m,r}$.
See Table \ref{sixbps} for the states in each representation.
\begin{table}[htb]
\caption{Fluctuation modes on a D3-brane wrapped on $X=0$.}\label{sixbps}
\centering
\begin{tabular}{cccccccccl}
\hline
\hline
rep. && $H$ & $J$ & $\ol J$ & $R_X$ & $R_Y$ & $R_Z$ & $A$ & range \\
\hline
${\cal B}^{{\rm D3}(X=0)}_{m,r}$
&        $R_-^m|X^*\rangle$ & $m+r-1$           & $0$ & $0$           & $-1$ & $m$ & $r$  & $0$ &  \\
&       $QR_-^m|X^*\rangle$ & $m+r-\frac{1}{2}$ & $\pm\frac{1}{2}$ & $0$ & $-\frac{1}{2}$ & $m+\frac{1}{2}$ & $r-\frac{1}{2}$ & $-1$ & $r\geq1$ \\
&     $Q^2R_-^m|X^*\rangle$ & $m+r$             & $0$ & $0$           & $0$ & $m+1$   & $r-1$ & $-2$ & $r\geq2$ \\
\hline
${\cal C}^{{\rm D3}(X=0)}_{m,r}$
&   $\ol QR_-^m|X^*\rangle$ & $m+r+\frac{1}{2}$ & $0$ & $+\frac{1}{2}$ & $-\frac{1}{2}$ & $m-\frac{1}{2}$ & $r+\frac{1}{2}$ & $+1$ & $m+r\geq0$ \\
&  $Q\ol QR_-^m|X^*\rangle$ & $m+r+1$             & $\pm\frac{1}{2}$ & $+\frac{1}{2}$ & $0$ & $m$ & $r$ & $0$ & $r\geq0$ \\
&$Q^2\ol QR_-^m|X^*\rangle$ & $m+r+\frac{3}{2}$ & $0$ & $+\frac{1}{2}$ & $+\frac{1}{2}$ & $m+\frac{1}{2}$ & $r-\frac{1}{2}$ & $-1$ & $r\geq1$ \\
\hline
\end{tabular}
\end{table}
The contributions of these representations to the index are as follows.
\begin{align}
\sum_m i({\cal B}^{{\rm D3}(X=0)}_{m,0})
&=(qu_x)^{-1}\frac{1}{1-qu_y},\nonumber\\
\sum_mi({\cal B}^{{\rm D3}(X=0)}_{m,1})
&=(qu_z)(qu_x)^{-1}\frac{1-q^{\frac{1}{2}}y^{\pm1}u_z^{-1}}{1-qu_y},\nonumber\\
\sum_mi({\cal B}^{{\rm D3}(X=0)}_{m,r>1})
&=(qu_z)^r(qu_x)^{-1}\frac{(1-q^{\frac{1}{2}}yu_z^{-1})(1-q^{\frac{1}{2}}y^{-1}u_z^{-1})}{1-qu_y},\nonumber\\
\sum_mi({\cal C}^{{\rm D3}(X=0)}_{m,r=-1})
&=-\frac{qu_y}{1-qu_y}+1,\nonumber\\
\sum_mi({\cal C}^{{\rm D3}(X=0)}_{m,r=0})
&=-(qu_z)\frac{1-q^{\frac{1}{2}}y^{\pm1}u_z^{-1}}{1-qu_y},\nonumber\\
\sum_mi({\cal C}^{{\rm D3}(X=0)}_{m,r\geq0})
&=-(qu_z)^{r+1}\frac{(1-q^{\frac{1}{2}}yu_z^{-1})(1-q^{\frac{1}{2}}y^{-1}u_z^{-1})}{1-qu_y}.
\label{ibcd3}
\end{align}

%%%%%%%%%%%%%%%%%%%%%%%%%%%%%%%%%%%%%%%%%%%%%%%%%%%%%%%%%%%%%%%%%%%%
\section{Current algebra}\label{affine.app}
We consider the current algebra of a simple Lie algebra $G$:
\begin{align}
[J_m^a,J_n^b]=if_{abc}J^c_{m+n}+km\delta_{a,b}\delta_{m+n,0},
\label{commj}
\end{align}
where $f_{abc}$ are the structure constants of $G$.
$k$ is the level, and we are interested in the basic representation of the algebra,
the Fock space constructed on a $G$-singlet ground state $|0\rangle$ with $k=1$.

The first few terms are
\begin{align}
\chi^{\wh G}=1+q\chi_\theta+q^2((\chi_\theta)^2_{\rm sym}-\chi_{2\theta}+\chi_\theta)+\cdots,
\label{threeterms}
\end{align}
where $(\chi_\theta)^2_{\rm sym}$ is the character of the symmetric product representation:
$(\chi_\theta)^2_{\rm sym}(x)=(\chi_\theta(x)^2+\chi_\theta(x^2))/2$.
The first term and the second term correspond to the ground state $|0\rangle$ and the first excited states
$J_{-1}^a|0\rangle$, respectively.
The third term corresponds to two types of grade $2$ states:
\begin{align}
J_{-1}^{\{a}J_{-1}^{b\}}|0\rangle,\quad
J_{-2}^a|0\rangle.
\label{grade2}
\end{align}
The former belong to the symmetric product of two copies of the adjoint representation,
which is decomposed into irreducible representations as
\begin{align}
(R_{\theta}\otimes R_{\theta})_{\rm sym}=R_{2\theta}+\cdots+R_0.
\end{align}
We denote the representation with the highest weight $w$ by $R_w$,
and $\theta$ is the highest weight of the adjoint representation.
With the commutation relation (\ref{commj})
we can easily show that states in $R_{2\theta}$ are null states.
By subtracting the null state contribution from the contribution of
(\ref{grade2}) we obtain the $q^2$ term in (\ref{threeterms}).
Higher order terms in $\chi^{\wh G}$ can be effectively obtained by
the free field realization.
\begin{align}
\chi^{\wh G}(q,x)=\frac{\sum_{p\in\Lambda_G}q^{\frac{p^2}{2}}x^p}{\prod_{n=1}^\infty(1-q^n)^r},\quad
r=\rank G.
\end{align}
$\Lambda_G$ is the root lattice of $G$.
$x$ collectively represents $r$ fugacities, and $x^p=\prod_{i=1}^rx_i^{p_i}$.
We show first few terms of $\chi^{\wh G}(q,1)$ for the seven types of $G$.
\begin{subequations}
\begin{align}
\chi^{\wh H_0}&=1,
\nonumber\\
\chi^{\wh H_1}
&=1+\chi^{H_1}_{\bm{3}}q
+(1+\chi^{H_1}_{\bm{3}})q^2
+(1+2\chi^{H_1}_{\bm{3}})q^3
+(2+2\chi^{H_1}_{\bm{3}}+\chi^{H_1}_{\bm{5}})q^4
+(2+4\chi^{H_1}_{\bm{3}}
\nonumber\\&
     +\chi^{H_1}_{\bm{5}})q^5
+(4+5\chi^{H_1}_{\bm{3}}+2\chi^{H_1}_{\bm{5}})q^6
+\cdots,\\
%%%%%%%%%%%%%%%%%%%%%%%%%%%%%%%%%
\chi^{\wh H_2}
&=1+\chi^{H_2}_{\bm{8}}q
+(1+2\chi^{H_2}_{\bm{8}})q^2
+(2+3\chi^{H_2}_{\bm{8}}+\chi^{H_2}_{\bm{10}}+\chi^{H_2}_{\ol{\bm{10}}})q^3
+(3+6\chi^{H_2}_{\bm{8}}+\chi^{H_2}_{\bm{10}}
\nonumber\\&
    +\chi^{H_2}_{\ol{\bm{10}}}+\chi^{H_2}_{\bm{27}})q^4
+(4+10\chi^{H_2}_{\bm{8}}+3\chi^{H_2}_{\bm{10}}+3\chi^{H_2}_{\ol{\bm{10}}}+2\chi^{H_2}_{\bm{27}})q^5
+(8+16\chi^{H_2}_{\bm{8}}+5\chi^{H_2}_{\bm{10}}
\nonumber\\&
    +5\chi^{H_2}_{\ol{\bm{10}}}+5\chi^{H_2}_{\bm{27}})q^6
+\cdots,\\
%%%%%%%%%%%%%%%%%%%%%%%%%%%%%%%%%%%
\chi^{\wh D_4}&=1+\chi^{D_4}_{\bm{28}}q
+(1+\chi^{D_4}_{\bm{28}}+\chi^{D_4}_{\bm{35}_V}+\chi^{D_4}_{\bm{35}_S}+\chi^{D_4}_{\bm{35}_C})q^2
+(1+4\chi^{D_4}_{\bm{28}}+\chi^{D_4}_{\bm{35}_V}+\chi^{D_4}_{\bm{35}_S}
\nonumber\\&
    +\chi^{D_4}_{\bm{35}_C}+\chi^{D_4}_{\bm{350}})q^3
+(4+5\chi^{D_4}_{\bm{28}}+3\chi^{D_4}_{\bm{35}_V}+3\chi^{D_4}_{\bm{35}_S}+3\chi^{D_4}_{\bm{35}_C}+\chi^{D_4}_{\bm{300}}+3\chi^{D_4}_{\bm{350}})q^4
\nonumber\\&
+(4+12\chi^{D_4}_{\bm{28}}+5\chi^{D_4}_{\bm{35}_V}+5\chi^{D_4}_{\bm{35}_S}+5\chi^{D_4}_{\bm{35}_C}+\chi^{D_4}_{\bm{300}}+7\chi^{D_4}_{\bm{350}}
      +\chi^{D_4}_{\bm{567}_V}+\chi^{D_4}_{\bm{567}_S}
\nonumber\\&
      +\chi^{D_4}_{\bm{567}_C})q^5
+(9+18\chi^{D_4}_{\bm{28}}+11\chi^{D_4}_{\bm{35}_V}
       +11\chi^{D_4}_{\bm{35}_S}+11\chi^{D_4}_{\bm{35}_C}
       +5\chi^{D_4}_{\bm{300}}+14\chi^{D_4}_{\bm{350}}
\nonumber\\&
      +2\chi^{D_4}_{\bm{567}_V}
      +2\chi^{D_4}_{\bm{567}_S}+2\chi^{D_4}_{\bm{567}_C}
      +\chi^{D_4}_{\bm{840}_V}+\chi^{D_4}_{\bm{840}_S}+\chi^{D_4}_{\bm{840}_C})q^6
+\cdots,\\
%%%%%%%%%%%%%%%%%%%%%%%%%%%%%%%%%%%
\chi^{\wh E_6}
&=1+\chi^{E_6}_{\bm{78}}q
+(1+\chi^{E_6}_{\bm{78}}+\chi^{E_6}_{\bm{650}})q^2
+(1+2\chi^{E_6}_{\bm{78}}+2\chi^{E_6}_{\bm{650}}+\chi^{E_6}_{\bm{2925}})q^3
+(2+4\chi^{E_6}_{\bm{78}}
\nonumber\\&
    +4\chi^{E_6}_{\bm{650}}+\chi^{E_6}_{\bm{2430}}+\chi^{E_6}_{\bm{2925}}+\chi^{E_6}_{\bm{5824}}+\chi^{E_6}_{\ol{\bm{5824}}})q^4
+(3+7\chi^{E_6}_{\bm{78}}+7\chi^{E_6}_{\bm{650}}+\chi^{E_6}_{\bm{2430}}
\nonumber\\&
    +4\chi^{E_6}_{\bm{2925}}
    +2\chi^{E_6}_{\bm{5824}}+2\chi^{E_6}_{\ol{\bm{5824}}}+\chi^{E_6}_{\bm{34749}})q^5
+(6+11\chi^{E_6}_{\bm{78}}+14\chi^{E_6}_{\bm{650}}+3\chi^{E_6}_{\bm{2430}}
\nonumber\\&
    +7\chi^{E_6}_{\bm{2925}}
+\chi^{E_6}_{\bm{3003}}+\chi^{E_6}_{\ol{\bm{3003}}}
+4\chi^{E_6}_{\bm{5824}}+4\chi^{E_6}_{\ol{\bm{5824}}}+3\chi^{E_6}_{\bm{34749}}+\chi^{E_6}_{\bm{70070}})q^6
+\cdots,\\
%%%%%%%%%%%%%%%%%%%%%%%%
\chi^{\wh E_7}
&=1+\chi^{E_7}_{\bm{133}}q+(1+\chi^{E_7}_{\bm{133}}+\chi^{E_7}_{\bm{1539}})q^2
+(1+2\chi^{E_7}_{\bm{133}}+\chi^{E_7}_{\bm{1463}}+\chi^{E_7}_{\bm{1539}}+\chi^{E_7}_{\bm{8645}})q^3
\nonumber\\&
+(2+3\chi^{E_7}_{\bm{133}}+\chi^{E_7}_{\bm{1463}}+3\chi^{E_7}_{\bm{1539}}+\chi^{E_7}_{\bm{7371}}+\chi^{E_7}_{\bm{8645}}+\chi^{E_7}_{\bm{40755}})q^4
+(2+6\chi^{E_7}_{\bm{133}}
\nonumber\\&
    +3\chi^{E_7}_{\bm{1463}}+4\chi^{E_7}_{\bm{1539}}+\chi^{E_7}_{\bm{7371}}+3\chi^{E_7}_{\bm{8645}}+2\chi^{E_7}_{\bm{40755}}+\chi^{E_7}_{\bm{152152}})q^5
+(5+8\chi^{E_7}_{\bm{133}}
\nonumber\\&
+4\chi^{E_7}_{\bm{1463}}+9\chi^{E_7}_{\bm{1539}}+3\chi^{E_7}_{\bm{7371}}
+5\chi^{E_7}_{\bm{8645}}+4\chi^{E_7}_{\bm{40755}}+\chi^{E_7}_{\bm{150822}}
+2\chi^{E_7}_{\bm{152152}}
\nonumber\\&
    +\chi^{E_7}_{\bm{365750}})q^6
+\cdots,\\
%%%%%%%%%%%%%%%%%%%%%%%%
\chi^{\wh E_8}
&=1
+\chi^{E_8}_{\bm{248}}q+q^2(1+\chi^{E_8}_{\bm{248}}+\chi^{E_8}_{\bm{3875}})q
+(1 + 2\chi^{E_8}_{\bm{248}}+\chi^{E_8}_{\bm{3875}}+\chi^{E_8}_{\bm{30380}})q^3
+(2
\nonumber\\&
    +3\chi^{E_8}_{\bm{248}}+2\chi^{E_8}_{\bm{3875}}+\chi^{E_8}_{\bm{27000}}+\chi^{E_8}_{\bm{30380}}+\chi^{E_8}_{\bm{147250}})q^4
+(2+5\chi^{E_8}_{\bm{248}}+3\chi^{E_8}_{\bm{3875}}
\nonumber\\&
    +\chi^{E_8}_{\bm{27000}}+3\chi^{E_8}_{\bm{30380}}+\chi^{E_8}_{\bm{147250}}+\chi^{E_8}_{\bm{779247}})q^5
+(4+7\chi^{E_8}_{\bm{248}}+6\chi^{E_8}_{\bm{3875}}+3\chi^{E_8}_{\bm{27000}}
\nonumber\\&
    +4\chi^{E_8}_{\bm{30380}}+2\chi^{E_8}_{\bm{147250}}+2\chi^{E_8}_{\bm{779247}}+\chi^{E_8}_{\bm{2450240}})q^6
+\cdots
\end{align}
\end{subequations}

%%%%%%%%%%%%%%%%%%%%%%%%%%%%%%%%%%%%%%%%%%%%%%%%%%%%%%%%%%
\section{Kaluza-Klein index}\label{largen.app}
The bulk contributions calculated by (\ref{bulkcont}) are shown below.
\begin{subequations}
\begin{align}
{\cal I}^{\rm KK}_{H_0}
&=1+u_z^{\frac{6}{5}}q^{\frac{6}{5}}
-u_z^{\frac{1}{5}}\chi^J_1q^{\frac{17}{10}}
+u_z^{-1}\chi^F_2q^2
+(u_z^{\frac{7}{10}}\chi^F_1+u_z^{-\frac{4}{5}})q^{\frac{11}{5}}
+2u_z^{\frac{12}{5}}q^{\frac{12}{5}}
\nonumber\\&
+(u_z^{\frac{6}{5}}-u_z^{-\frac{3}{10}}\chi^F_1)\chi^J_1q^{\frac{27}{10}}
-2u_z^{\frac{7}{5}}\chi^J_1q^{\frac{29}{10}}
+(-1+u_z^{-\frac{3}{2}}\chi^F_3-\chi^F_2)q^3
+\cdots,\\
%%%%%%%%%%%%%%%%%%%%%%%%%%%%%%%%%%%%%%%%%%%%%%%%%%%
{\cal I}_{H_1}^{\rm KK}
&=1
+u_z^{\frac{4}{3}}q^{\frac{4}{3}}
-u_z^{\frac{1}{3}}\chi^J_1q^{\frac{11}{6}}
+(u_z^{-1}\chi^F_2+u_z^{-1}\chi^{H_1}_{\bm{3}})q^2
+(u_z^{-\frac{2}{3}}+\chi^F_1u_z^{\frac{5}{6}})q^{\frac{7}{3}}
+2u_z^{\frac{8}{3}}q^{\frac{8}{3}}
\nonumber\\&
+(-u_z^{-\frac{1}{6}}\chi^F_1+u_z^{\frac{4}{3}})\chi^J_1q^{\frac{17}{6}}
+(-1-\chi^F_2+u_z^{-\frac{3}{2}}\chi^F_3-\chi^{H_1}_{\bm{3}}
    +u_z^{-\frac{3}{2}}\chi^F_1\chi^{H_1}_{\bm{3}})q^3
\nonumber\\&
-2u_z^{\frac{5}{3}}\chi^J_1q^{\frac{19}{6}}
+(u_z^{-\frac{7}{6}}\chi^F_1-u_z^{\frac{1}{3}}+2u_z^{\frac{1}{3}}\chi^F_2
   -uz^{\frac{11}{6}}\chi^F_1-u_z^{\frac{1}{3}}\chi^J_2+u_z^{\frac{1}{3}}\chi^{H_1}_{\bm{3}})q^{\frac{10}{3}}
\nonumber\\&
+(u_z^{-1} +u_z^{-1}\chi^F_2+u_z^{-1}\chi^{H_1}_{\bm{3}})\chi^J_1q^{\frac{7}{2}}
+(3u_z^{\frac{2}{3}}+2u_z^{\frac{13}{6}}\chi^F_1)q^{\frac{11}{3}}
+\cdots,\\
%%%%%%%%%%%%%%%%%%%%%%%%%%%%%%%%%%%%%%%%%%%%%%%%%%%
{\cal I}_{H_2}^{\rm KK}
&=1
+u_z^{\frac{3}{2}}q^{\frac{3}{2}}
+(u_z^{-1}\chi^F_2-u_z^{\frac{1}{2}}\chi^J_1+u_z^{-1}\chi^{H_2}_{\bm{8}})q^2
+(u_z^{-\frac{1}{2}}+u_z\chi^F_1)q^{\frac{5}{2}}
+(-1-\chi^F_2
\nonumber\\&
             +u_z^{\frac{3}{2}}\chi^F_3+2u_z^3-\chi^F_1\chi^J_1
    +u_z^{\frac{3}{2}}\chi^J_1-\chi^{H_2}_{\bm{8}}+u_z^{-\frac{3}{2}}\chi^F_1\chi^{H_2}_{\bm{8}})q^3
+(u_z^{-1}\chi^F_1-u_z^{\frac{1}{2}}
       +2u_z^{\frac{1}{2}}\chi^F_2
\nonumber\\&
-u_z^2\chi^F_1+u_z^{-1}\chi^J_1
   +u_z^{-1}\chi^F_2\chi^J_1
   -2u_z^2\chi^J_1-u_z^{\frac{1}{2}}\chi^J_2
   +u_z^{\frac{1}{2}}\chi^{H_2}_{\bm{8}}
   +u_z^{-1}\chi^{H_2}_{\bm{8}}\chi^J_1)q^{\frac{7}{2}}
\nonumber\\&
+(2u_z^{-2}+2u_z^{-2}\chi^F_4-u_z^{-\frac{1}{2}}\chi^F_1
      -u_z^{-\frac{1}{2}}\chi^F_3+4u_z+2u_z^{\frac{5}{2}}\chi^F_1
      +u_z^{-\frac{1}{2}}\chi^J_1
      -2u_z^{-\frac{1}{2}}\chi^F_2\chi^J_1
\nonumber\\&
      +2u_z\chi^F_1\chi^J_1+u_z^{-2}\chi^{H_2}_{\bm{8}}+2u_z^{-2}\chi^F_2\chi^{H_2}_{\bm{8}}
      -u_z^{-\frac{1}{2}}\chi^F_1\chi^{H_2}_{\bm{8}}
      -u_z^{-\frac{1}{2}}\chi^J_1\chi^{H_2}_{\bm{8}}
      +u_z^2\chi^{H_2}_{\bm{27}})q^4
\nonumber\\&
+(-\chi^F_1+3\chi^F_3+2u_z^{-\frac{3}{2}}\chi^F_2
     -2u_z^{\frac{3}{2}}-2u_z^{\frac{3}{2}}\chi^F_2+3u_z^{\frac{9}{2}}
     -3\chi^J_1-\chi^F_2\chi^J_1
     +u_z^{-\frac{3}{2}}\chi^F_1\chi^J_1
\nonumber\\&
     +u_z^{-\frac{3}{2}}\chi^F_3\chi^J_1
     -3u_z^{\frac{3}{2}}\chi^F_1\chi^J_1+2u_z^3\chi^J_1-\chi^F_1\chi^J_2
     +u_z^{\frac{3}{2}}\chi^J_2
     +2\chi^F_1\chi^{H_2}_{\bm{8}}
     +u_z^{-\frac{3}{2}}\chi^{H_2}_{\bm{8}}
\nonumber\\&
     -u_z^{\frac{3}{2}}\chi^{H_2}_{\bm{8}}-\chi^J_1\chi^{H_2}_{\bm{8}}
     +u_z^{-\frac{3}{2}}\chi^F_1\chi^J_1\chi^{H_2}_{\bm{8}})q^{\frac{9}{2}}
+\cdots,\\
%%%%%%%%%%%%%%%%%%%%%%%%%%%%%%%%%%%%%%%%%%%%%%%%%%%
{\cal I}_{D_4}^{\rm KK}
=&
1
+(u_z^2+\chi_2^F u_z^{-1} + u_z^{-1} \chi^{D_4}_{\bm{28}})q^2
-u_z \chi_1^Jq^{\frac{5}{2}}
+(-\chi^F_2 + \chi^F_1u_z^{\frac{3}{2}} +\chi^F_3 u_z^{-\frac{3}{2}}
       - \chi^{D_4}_{\bm{28}}
\nonumber\\&
       + \chi^F_1 u_z^{-\frac{3}{2}} \chi^{D_4}_{\bm{28}})q^3
+(u_z^2\chi^J_1 - \chi^F_1 u_z^{\frac{1}{2}}\chi^J_1
   + u_z^{-1}\chi^J_1 + \chi^F_2u_z^{-1}\chi^J_1 + u_z^{-1}\chi^J_1 \chi^{D_4}_{\bm{28}})q^{\frac{7}{2}}
\nonumber\\&
+(2 u_z^4 - \chi^F_1 u_z^{\frac{5}{2}} + 2 \chi^F_2 u_z
  -\chi^F_3u_z^{-\frac{1}{2}} + 2 u_z^{-2}
   + 2 \chi^F_4 u_z^{-2} - u_z\chi^J_2
   + (u_z-\chi^F_1 u_z^{-\frac{1}{2}}
\nonumber\\&
            + 2 \chi^F_2 u_z^{-2}) \chi^{D_4}_{\bm{28}}
     + u_z^{-2} \chi^{D_4}_{\bm{300}}
      + u_z^{-2} \chi^{D_4}_{\bm{35}_v}
       + u_z^{-2} \chi^{D_4}_{\bm{35}_s}
        + u_z^{-2} \chi^{D_4}_{\bm{35}_c})q^4
+(2u_z^{-\frac{5}{2}}\chi^F_1
\nonumber\\&
  +u_z^{-\frac{5}{2}}\chi^F_3
  +2u_z^{-\frac{5}{2}}\chi^F_5
  -u_z^{-1}
  -u_z^{-1}\chi^F_2
  -2u_z^{-1}\chi^F_4
  +3u_z^{\frac{1}{2}}\chi^F_3
  +u_z^2
  -2u_z^2\chi^F_2
\nonumber\\&
  +2u_z^{\frac{7}{2}}\chi^F_1
  +u_z^{-1}\chi^J_2
  +u_z^{-1}\chi^F_2\chi^J_2
  -u_z^{\frac{1}{2}}\chi^F_1\chi^J_2
  +u_z^2\chi^J_2
  +2u_z^{-\frac{5}{2}}\chi^F_1\chi^{D_4}_{\bm{28}}
\nonumber\\&
  +3u_z^{-\frac{5}{2}}\chi^F_3\chi^{D_4}_{\bm{28}}
  -u_z^{-1}\chi^{D_4}_{\bm{28}}
  -3u_z^{-1}\chi^F_2\chi^{D_4}_{\bm{28}}
  +2u_z^{-\frac{1}{2}}\chi^F_1\chi^{D_4}_{\bm{28}}
  -u_z^2\chi^{D_4}_{\bm{28}}
  +u_z^{-1}\chi^J_2\chi^{D_4}_{\bm{28}}
\nonumber\\&
  +u_z^{-\frac{5}{2}}\chi^F_1(\chi^{D_4}_{\bm{35}_v}+\chi^{D_4}_{\bm{35}_s}+\chi^{D_4}_{\bm{35}_c})
  -u_z^{-1}(\chi^{D_4}_{\bm{35}_v}+\chi^{D_4}_{\bm{35}_s}+\chi^{D_4}_{\bm{35}_c})
  +u_z^{-\frac{5}{2}}\chi^F_1\chi^{D_4}_{\bm{300}}
\nonumber\\&
  -u_z^{-1}\chi^{D_4}_{\bm{300}}
  +u_z^{-\frac{5}{2}}\chi^F_1\chi^{D_4}_{\bm{350}}
  -u_z^{-1}\chi^{D_4}_{\bm{350}})q^5
+\cdots,
\label{ikkd4}\\
%%%%%%%%%%%%%%%%%%%%%%%%%%%%%%%%%%%%%%%%%%%%%%%%%%%
{\cal I}_{E_6}^{\rm KK}
&=1
+(u_z^{-1}\chi^F_2 + u_z^{-1}\chi^{E_6}_{\bm{78}})q^2
+(-1-\chi^F_2 +u_z^{-\frac{3}{2}}\chi^F_3
   +u_z^3-\chi^{E_6}_{\bm{78}}+u_z^{-\frac{3}{2}}\chi^F_1\chi^{E_6}_{\bm{78}})q^3
\nonumber\\&
+(u_z^{-1}+u_z^{-1}\chi^F_2-u_z^2+u_z^{-1}\chi^{E_6}_{\bm{78}})\chi^J_1q^{\frac{7}{2}}
+(2u_z^{-2}+2u_z^{-2}\chi^F_4-u_z^{-\frac{1}{2}}\chi^F_1-u_z^{-\frac{1}{2}}\chi^F_3
\nonumber\\&
    +2u_z
    +uz^{\frac{5}{2}}\chi^F_1
    +2u_z^{-2}\chi^F_2\chi^{E_6}_{\bm{78}}-u_z^{-\frac{1}{2}}\chi^F_1\chi^{E_6}_{\bm{78}}
    +u_z^{-2}\chi^{E_6}_{\bm{650}}+u_z^{-2}\chi^{E_6}_{\bm{2430}})q^4
+(-2
\nonumber\\&
    -\chi^F_2
    +u_z^{-\frac{3}{2}}\chi^F_1+u_z^{-\frac{3}{2}}\chi^F_3-uz^{\frac{3}{2}}\chi^F_1
    +u_z^3-\chi_{78}+u_z^{-\frac{3}{2}}\chi^F_1\chi^{E_6}_{\bm{78}})\chi^J_1q^{\frac{9}{2}}
+(2u_z^{-\frac{5}{2}}\chi^F_1
\nonumber\\&
    +uz^{-\frac{5}{2}}\chi^F_3+2u_z^{-\frac{5}{2}}\chi^F_5-u_z^{-1}
    -3u_z^{-1}\chi^F_2-2u_z^{-1}\chi^F_4
    +2u_z^{\frac{1}{2}}\chi^F_1-u_z^2+2u_z^2\chi^F_2
\nonumber\\&
    -u_z^{\frac{7}{2}}\chi^F_1
    +u_z^{-1}\chi^J_2+u_z^{-1}\chi^F_2\chi^J_2-u_z^2\chi^J_2
    +2u_z^{-\frac{5}{2}}\chi^F_1\chi^{E_6}_{\bm{78}}+3u_z^{-\frac{5}{2}}\chi^F_3\chi^{E_6}_{\bm{78}}-2u_z\chi^{E_6}_{\bm{78}}
\nonumber\\&
    -3u_z\chi^F_2\chi^{E_6}_{\bm{78}}+u_z^2\chi^{E_6}_{\bm{78}}+u_z\chi^J_2\chi^{E_6}_{\bm{78}}
    +u_z^{-\frac{5}{2}}\chi^F_1\chi^{E_6}_{\bm{650}}
    -u_z^{-1}\chi^{E_6}_{\bm{650}}+u_z^{-\frac{5}{2}}\chi^F_1\chi^{E_6}_{\bm{2430}}
\nonumber\\&
    -u_z^{-1}\chi^{E_6}_{\bm{2430}}
    +u_z^{-\frac{5}{2}}\chi^F_1\chi^{E_6}_{\bm{2925}}-u_z^{-1}\chi^{E_6}_{\bm{2925}})q^5
+\cdots,\\
%%%%%%%%%%%%%%%%%%%%%%%%%%%%%%%%%%%%%%%%%%%%%%%%%%%
{\cal I}_{E_7}^{\rm KK}
&=1
+(u_z^{-1}\chi^F_2
    +u_z^{-1}\chi^{E_7}_{\bm{133}})q^2
+(-1
     -\chi^F_2
     +u_z^{-\frac{3}{2}}\chi^F_3
     -\chi^{E_7}_{\bm{133}}
     +u_z^{-\frac{3}{2}}\chi^F_1\chi^{E_7}_{\bm{133}})q^3
\nonumber\\&
+(u_z^{-1}
    +u_z^{-1}\chi^F_2
    +u_z^{-1}\chi^{E_7}_{\bm{133}})\chi^J_1q^{\frac{7}{2}}
+(2u_z^{-2}
    +2u_z^{-2}\chi^F_4
    -u_z^{-\frac{1}{2}}\chi^F_1
    -u_z^{-\frac{1}{2}}\chi^F_3
\nonumber\\&
    +u_z
    +u_z^4
    +2u_z^{-2}\chi^F_2\chi^{E_7}_{\bm{133}}
    -u_z^{-\frac{1}{2}}\chi^F_1\chi^{E_7}_{\bm{133}}
    +u_z^{-2}\chi^{E_7}_{\bm{1539}}
    +u_z^{-2}\chi^{E_7}_{\bm{7371}})q^4
+(-2
\nonumber\\&
    -\chi^F_2
    +u_z^{-\frac{3}{2}}\chi^F_1
    +u_z^{-\frac{3}{2}}\chi^F_3
    -u_z^3
    -\chi^{E_7}_{\bm{133}}
    +u_z^{-\frac{3}{2}}\chi^F_1\chi^{E_7}_{\bm{133}})\chi^J_1q^{\frac{9}{2}}
+(2u_z^{-\frac{5}{2}}\chi^F_1
    +u_z^{-\frac{5}{2}}\chi^F_3
\nonumber\\&
    +2u_z^{-\frac{5}{2}}\chi^F_5
    -u_z^{-1}
    -3u_z^{-1}\chi^F_2
    -2u_z^{-1}\chi^F_4
    +u_z^{\frac{1}{2}}\chi^F_1
    +u_z^2
    +u_z^{\frac{7}{2}}\chi^F_1
    +u_z^{-1}\chi^J_2
\nonumber\\&
    +u_z^{-1}\chi^F_2\chi^J_2
    +2u_z^{-\frac{5}{2}}\chi^F_1\chi^{E_7}_{\bm{133}}
    +3u_z^{-\frac{5}{2}}\chi^F_3\chi^{E_7}_{\bm{133}}
    -2u_z^{-1}\chi^{E_7}_{\bm{133}}
    -3u_z^{-1}\chi^F_2\chi^{E_7}_{\bm{133}}
\nonumber\\&
    +u_z^{-1}\chi^J_2\chi^{E_7}_{\bm{133}}
    +u_z^{-\frac{5}{2}}\chi^F_1\chi^{E_7}_{\bm{1539}}
    -u_z^{-1}\chi^{E_7}_{\bm{1539}}
    +u_z^{-\frac{5}{2}}\chi^F_1\chi^{E_7}_{\bm{7371}}
    -u_z^{-1}\chi^{E_7}_{\bm{7371}}
\nonumber\\&
    +u_z^{-\frac{5}{2}}\chi^F_1\chi^{E_7}_{\bm{8645}}
    -u_z^{-1}\chi^{E_7}_{\bm{8645}})q^5
+\cdots,\\
%%%%%%%%%%%%%%%%%%%%%%%%%%%%%%%%%%%%%%%%%%%%%%%%%%%
{\cal I}_{E_8}^{\rm KK}
&=1
+(u_z^{-1}\chi^F_2+u_z^{-1}\chi^{E_8}_{\bm{248}})q^2
+(-1-\chi^F_2+u_z^{-\frac{3}{2}}\chi^F_3-\chi^{E_8}_{\bm{248}}+u_z^{-\frac{3}{2}}\chi^F_1\chi^{E_8}_{\bm{248}})q^3
\nonumber\\&
+(u_z^{-1}+u_z^{-1}\chi^F_2+u_z^{-1}\chi^{E_8}_{\bm{248}})\chi^J_1q^{\frac{7}{2}}
+(2u_z^{-2}+2u_z^{-2}\chi^F_4-u_z^{-\frac{1}{2}}\chi^F_1-u_z^{-\frac{1}{2}}\chi^F_3
\nonumber\\&
    +u_z+2u_z^{-2}\chi^F_2\chi^{E_8}_{\bm{248}}-u_z^{-\frac{1}{2}}\chi^F_1\chi^{E_8}_{\bm{248}}
    +u_z^{-2}\chi^{E_8}_{\bm{3875}}+u_z^{-2}\chi^{E_8}_{\bm{27000}})q^4
+(-2-\chi^F_2
\nonumber\\&
    +u_z^{-\frac{3}{2}}\chi^F_1
    +u_z^{-\frac{3}{2}}\chi^F_3-\chi^{E_8}_{\bm{248}}+u_z^{-\frac{3}{2}}\chi^F_1\chi^{E_8}_{\bm{248}})\chi^J_1q^{\frac{9}{2}}
+(2u_z^{-\frac{5}{2}}\chi^F_1+u_z^{-\frac{5}{2}}\chi^F_3
\nonumber\\&
    +2u_z^{-\frac{5}{2}}\chi^F_5
    -u_z^{-1}-3u_z^{-1}\chi^F_2
    -2u_z^{-1}\chi^F_4
    +u_z^{\frac{1}{2}}\chi^F_1
    +u_z^{-1}\chi^J_2
    +u_z^{-1}\chi^F_2\chi^J_2
\nonumber\\&
    +2u_z^{-\frac{5}{2}}\chi^F_1\chi^{E_8}_{\bm{248}}
    +3u_z^{-\frac{5}{2}}\chi^F_3\chi^{E_8}_{\bm{248}}
    -2u_z^{-1}\chi^{E_8}_{\bm{248}}
    -3u_z^{-1}\chi^F_2\chi^{E_8}_{\bm{248}}
    +u_z^{-1}\chi^{E_8}_{\bm{248}}\chi^J_2
\nonumber\\&
    +u_z^{-\frac{5}{2}}\chi^F_1\chi^{E_8}_{\bm{3875}}
    -u_z^{-1}\chi^{E_8}_{\bm{3875}}
    +u_z^{-\frac{5}{2}}\chi^F_1\chi^{E_8}_{\bm{27000}}
    -u_z^{-1}\chi^{E_8}_{\bm{27000}}
    +u_z^{-\frac{5}{2}}\chi^F_1\chi^{E_8}_{\bm{30380}}
\nonumber\\&
    -u_z^{-1}\chi^{E_8}_{\bm{30380}})q^5
+\cdots.
\end{align}
\end{subequations}

%%%%%%%%%%%%%%%%%%%%%%%%%%%%%%%%%%%%%%%%%%%%%%%%%%%%%%%%%%%%%

\end{document}